\documentclass[]{pasj01}
\draft
\usepackage{natbib}
\usepackage{color}
\usepackage{ulem}

\begin{document} 
\Received{}
\Accepted{}

\title{Properties of environment around AGN and luminous galaxy pairs through HSC wide survey}


\author{Yuji \textsc{Shirasaki}\altaffilmark{1,2}}%
\author{Masayuki \textsc{Akiyama}\altaffilmark{3}}%
\author{Yoshiki \textsc{Toba}\altaffilmark{4,5,6}}%
\author{Wanqiu \textsc{He}\altaffilmark{3}}%
\author{Tomotsugu \textsc{Goto}\altaffilmark{7}}%

\altaffiltext{1}{National Astronomical Observatory of Japan,
2-21-1 Osawa, Mitaka, Tokyo 181-8588, Japan}
\email{yuji.shirasaki@nao.ac.jp}

\altaffiltext{2}{Department of Astronomical Science, School 
of Physical Sciences, The Graduate University for 
Advanced Studies, SOKENDAI,
2-21-1 Osawa, Mitaka, Tokyo 181-8588, Japan}

\altaffiltext{3}{Astronomical Institute, Tohoku University,
6-3 Aramaki, Aoba-ku Sendai, Miyagi 980-8578, Japan}

\altaffiltext{4} {Department of Astronomy, Kyoto University, 
Kitashirakawa-Oiwake-cho, Sakyo-ku, Kyoto 606-8502, Japan}
 
\altaffiltext{5}{Academia Sinica Institute of Astronomy and
Astrophysics, 11F of Astronomy-Mathematics Building,
AS/NTU, No.1, Section 4, Roosevelt Road, Taipei 10617, Taiwan}

\altaffiltext{6}{Research Center for Space and Cosmic Evolution,
Ehime University, 
10-13 Dogo-Himata, Matsuyama, Ehime 790-8577, Japan}

\altaffiltext{7} {Institute of Astronomy and Department of Physics, National
Tsing Hua University, Hsinchu 30013, Taiwan}



\maketitle

\begin{abstract}
We investigated the properties of AGN 
environments, 
particularly 
environments
where the association of luminous galaxies (LGs) is found 
within 4~Mpc from AGNs with redshifts of 0.8 -- 1.1.
For comparison, three additional AGN 
environments, 
(namely, AGNs of all types, 
type~1 AGNs with X-ray and/or radio detection, and type~2 AGNs) and an
environment 
of blue $M_{*}$, 
characteristic luminosity of the Schechter function, 
galaxies were investigated.
The cross-correlation function with the surrounding galaxies was measured
and compared between the AGN and blue galaxy samples.
We also compared the distributions of color, absolute magnitude, and stellar mass of
the galaxies around such target objects. 
The properties of clusters detected using surrounding galaxies
selected based on a photometric redshift were examined and compared
for different samples.
The target AGNs were drawn from the Million Quasars (MILLIQUAS) catalog,
and the blue galaxies were drawn from six redshift survey catalogs 
(SDSS, WiggleZ, DEEP2, VVDS, VIPERS, and PRIMUS).
The galaxies used as a measure of the environment around the targets
are drawn from S18a internal data released
by the Hyper Suprime-Cam Subaru Strategic Program (HSC-SSP).
We found that, among the five AGN and blue galaxy samples considered, 
the environment of AGN-LG pairs is the most enriched with 
luminous galaxies. 
We also found an enhancement in the number of mass-selected clusters 
in the AGN-LG pair sample against those in the other samples.
The results obtained in this study indicate that existence of multiple 
clusters is the major driver in the association of AGNs and LGs, rather than
a single large-mass dark matter halo hosting the AGN.
\end{abstract}

\section{Introduction}
\label{sec:introduction}

The ubiquity of super massive black holes (SMBHs) at the centers of
galaxies has been recognized through the observation of nearby galaxies
\citep{Richstone+98,Kormendy+13}.
Various processes are believed to be relevant to the feeding of SMBHs,
including the following: 
a secular evolution caused by gravitational instability inside a galaxy
\citep{Kormendy+04}, 
a minor/major merger~\citep{Sanders+88,DiMatteo+05,Hopkins+08}, 
a quiescent accretion of hot halo gas~\citep{Keres+09,Fanidakis+13}, the
ram pressure~\citep{Bosch+08}, and
tidal force feeding.

Low and intermediate luminosity AGNs are thought to be mostly caused 
by a secular evolution, and the most luminous AGNs, i.e., QSOs, are
triggered by major mergers~\citep[e.g.][]{Treister+12,Menci+14}.
Several observations, however, have indicated that the major mergers are 
not a dominant mechanism for the triggering of a QSO~\citep[e.g.][]{Villforth+17}.
The main driving mechanism of QSO activity remains
open to debate.

One of the powerful ways to distinguish among the possible mechanisms
is the environmental analysis of the AGN~\citep[e.g.][]{Coil+07,Coil+09,
    Hickox+09,Hickox+11,Krumpe+12,
    Krumpe+18,Komiya+13,Ikeda+15,Shirasaki+18,He+18}.
Mergers are likely to occur in a high density region except for a
cluster core, where the relative velocity is too high to merge or interact 
with other galaxies and trigger a gas accretion.
The quiescent accretion of hot halo gas may be found in a cluster core, 
whereas ram pressure feeding occurs in the in-fall region of a cluster and
between clusters that are in the process of merging.
Thus, knowing the positional relationship between AGNs and high density 
regions provides a hint to solve the present problem.

The Hyper Suprime-Cam Subaru Strategic Program (HSC-SSP) is a multi-band
imaging survey conducted using the 
HSC~\citep{Miyazaki+12,Miyazaki+18,Komiyama+18,Kawanomoto+18,Furusawa+18}.
The dataset of the wide layer covers 1,400 deg$^2$, and the limiting
magnitude is as deep as $r \sim$ 26.
Thus, it provides a powerful tool to investigate the environment of 
an AGN with unprecedented statistics.
Using the first-year dataset of HSC-SSP, 
\citet{Shirasaki+18} measured the clustering of galaxies around AGNs, and
found that luminous galaxies are strongly clustered around them.
Their results indicate that the cross-correlation length increases from 
7 $h^{-1}$Mpc at approximately $M_{\lambda 310} = -19$ mag to $>$ 10 $h^{-1}$Mpc
beyond $M_{\lambda 310} = -20$ mag, where $M_{\lambda 310}$ represents the absolute magnitude
measured at the rest frame wavelength of 310~nm.
At approximately $M_{\lambda 310} = -22$ mag, it reaches 30 $h^{-1}$Mpc, which
is too large to be attributed solely to the mass of the host dark matter halo;
the expected number density of dark matter haloes clustered
at the same level of those luminous galaxies is too low to produce
the observed number of luminous galaxies.
Thus, the large clustering of luminous galaxies should be attributed
in part, or mostly, to other properties related with their environment.

\citet{Shirasaki+18} also 
showed that the luminosity function
measured around AGNs can be described using a smaller 
(brighter)
characteristic luminosity, 
$M_{*}$,
parameter when fitted with the Schechter function~\citep{Schechter+76}.
This indicates that the mass assembly of galaxies rapidly
progress around some of the AGNs and, as a result, AGNs are more
likely to be associated with luminous galaxies.
As the strong cross-correlation between AGNs and luminous galaxies
extends over the $\sim$10~Mpc scale, the mechanism should be related
to the activity in a large-scale structure, such as a cluster-cluster
interaction, enhanced galaxy merger, or gas inflow at a saddle point in the 
filament, among other possibilities.
To understand what mechanism is relevant to the simultaneous
occurrence of an AGN activity and the evolution of galaxies around it,
it is crucial to investigate the properties of the environment
of the AGN, particularly an AGN with an association of luminous galaxies.

For this reason, we investigated the environmental properties around
AGNs associated with luminous galaxies (LGs) within a distance
of 4~Mpc (namely, AGNs paired with LGs, which are hereinafter referred to as AGN-LG 
pairs) by comparing with the environment
of four different target objects: blue galaxies,
AGNs as a whole,
type~1 AGNs through X-ray and/or radio detection (AGN type~1 XR), and type~2 AGNs.
To determine the distance scale for selecting AGN-LG pairs 
we examined the clustering of galaxies around the pairs for three different
scales, 2~Mpc, 4~Mpc and 8~Mpc, and found that the significance of the
excess of clustering against
the whole AGN sample
was the largest for the sample selected with 4~Mpc scale.
Thus we decided to use 4~Mpc for selecting AGN-LG pairs
for this study.
The blue galaxies are used as a proxy to the environment of
an ordinary galaxy.

Because different clustering properties
have been reported for different types of AGNs 
by numerous authors~\mbox{\citep[e.g.][]{Hickox+09,Allevato+14,Mendez+16}},
we also carried out a comparison between different types of AGNs.
In those studies, AGNs selected by X-ray and radio observations
shows larger clustering compared to the other types of AGNs.
Thus the AGN type~1 XR sample is used as a representative
of AGNs in higher overdensity environments.
The clustering of LGs around the AGNs increases at higher redshifts~\citep{Shirasaki+18}, 
whereas the availability of galaxies measured for their redshift is
limited to a redshift of $\sim$1.1 
for the galaxy redshift catalogs 
used in this work:
SDSS DR14~\citep{Abolfathi+18}, 
WiggleZ final~\citep{Drinkwater+18}, 
DEEP2 DR4~\citep{Newman+13}, 
VVDS~\citep{Fevre+13}, 
VIPERS~\citep{Scodeggio+18}, 
and PRIMUS~\citep{Coil+11,Cool+13}.
Thus, the target redshift was set to 0.8--1.1, where
the sample size
for AGN-LG pairs becomes maximal.
From the results obtained through the comparisons,
we aim to determine what type of mechanism
has an effect on the AGN activity and the formation of 
LGs around the AGN.
Throughout this paper, we assume a cosmology with $\Omega_{m} = 0.3$, 
$\Omega_{\lambda} = 0.7$, $h = 0.7$, and $\sigma_{8} = 0.8$.
All magnitudes are given in the AB system.
All distances are measured in comoving coordinates.
The correlation length is presented in unit of $h^{-1}$Mpc to
facilitate a comparison with other measurements.

\section{Datasets}

\subsection{Galaxies as a measure of the environment}
\label{sec:hsc_source}
As a measure of the environment of the AGNs and blue galaxies considered, we used 
photometric galaxies derived from the HSC-SSP survey.
The internal release of the S18a wide layer dataset was used in this analysis.
The observed locations and effective area of the S18a wide dataset
are summarized in Table~\ref{tab:survey_param}.
The typical depths of the observation are 26.6, 26.2, 26.2, 25.3, and 24.5 for the
$g$, $r$, $i$, $z$, and $y$ bands, respectively.
The details of the survey itself are described in \citet{Aihara+17},
and the content of the S18a dataset is provided in \citet{Aihara+19}.

The S18a dataset was analyzed through the HSC pipeline
(version 6.5.1/6.5.3/6.6) developed by the HSC software team~\citep{Bosch+18}
using codes from the Large Synoptic
Survey Telescope (LSST) software pipeline~\citep{Ivezic+08,Axelrod+10,Juric+15}.
Photometric and astrometric calibrations were conducted based on
data obtained from the Panoramic Survey Telescope and
Rapid Response System (Pan-STARRS) 1 imaging 
survey~\citep{Magnier+13,Schlafly+12,Tonry+12}.

\begin{table}
  \tbl{Summary of the survey area}{%
  \begin{tabular}{cccr}
      \hline
Field name & Approx. center coordinates & $S^{a}$   \\
           &                    &  deg$^{2}$ \\
      \hline
WIDE12H/GAMA15H & 13$^{h}$10$^{m}$ $+$00$^{\circ}$00' &  263.1 \\
VVDS            & 23$^{h}$20$^{m}$ $+$02$^{\circ}$00' &  248.4  \\
GAMA09H         & 09$^{h}$35$^{m}$ $+$02$^{\circ}$00' &  196.9 \\
XMM-LSS         & 02$^{h}$15$^{m}$ $-$01$^{\circ}$00' &  132.5 \\
HECTOMAP        & 15$^{h}$00$^{m}$ $+$43$^{\circ}$30' &   98.4  \\
WIDE01H         & 01$^{h}$15$^{m}$ $+$01$^{\circ}$00' &   27.9  \\
AEGIS           & 14$^{h}$17$^{m}$ $+$52$^{\circ}$30' &    2.1  \\
\hline
Total           &                                     &  969.3 \\
      \hline
  \end{tabular}}\label{tab:survey_param}
  \begin{tabnote}
    $^{a}$ Effective area of each survey field of S18a internal release.
  \end{tabnote}
\end{table}

The photometric magnitude used in this work is a CModel magnitude.
The galactic reddening was corrected according to the dust maps derived
by \citet{Schlegel+98}.
There is a known issue regarding the CModel magnitude that, for some objects, the CModel magnitude has a significantly large deviation
from the other magnitude measurements, such as the magnitude in the aperture.
We checked the effect of such errors on our analysis, and found
that it is negligibly small.
The HSC sources satisfying the criteria summarized in 
table~\ref{tab:criteria} were selected.
The criteria were tested for all four $griz$-bands.
Because the observations in the $y$-band are shallower than those in the other bands,
the detection in the $y$-band was not required to avoid bias against
redder galaxies.
In addition to these selection criteria, other selections such 
as those based on the absolute magnitude space and photometric redshift
(photo-$z$)
were applied.
These additional selections will be described in the analysis method and
results section.

\begin{table}
\tbl{Summary of selection criteria of HSC sources}{
\begin{tabular}{lll}
\hline
column name & constraint & explanation of the column \\
\hline
\texttt{[girz]\_pixelflags\_edge}               & \texttt{IS NOT TRUE} & 
  Source is outside usable exposure region \\
\texttt{[griz]\_pixelflags\_saturatedcenter}    & \texttt{IS NOT TRUE} &  
  Saturated pixel in the Source center \\
\texttt{[griz]\_pixelflags\_bad}                & \texttt{IS NOT TRUE} &  
  Bad pixel in the Source footprint \\
\texttt{[griz]\_cmodel\_flag}                   & \texttt{IS NOT TRUE} &  
  cmodel fit failed\\
\texttt{[griz]\_cmodel\_mag}                    & \texttt{IS NOT NULL} &  
  cmodel magnitude \\
\texttt{[griz]\_cmodel\_mag}                    & \texttt{BETWEEN 1 AND 90} &  \\
\texttt{[griz]\_cmodel\_magsigma}               & \texttt{IS NOT NULL}      &
  uncertainty of cmodel magnitude \\
\texttt{[griz]\_cmodel\_magsigma}               & \texttt{BETWEEN 0 AND 0.2} &  \\
\texttt{i\_mask\_s18a\_bright\_objectcenter}    & \texttt{IS NOT TRUE}    &  
  Source center is close to \texttt{BRIGHT\_OBJECT} pixels \\
\texttt{isprimary}                              & \texttt{IS TRUE}    &  
  true if this is a primary data of this object \\
\hline
\end{tabular}}\label{tab:criteria}
\begin{tabnote}
\texttt{[griz]} in the column name means that any corresponding columns of
four HSC bands ($g$,$r$,$i$, and $z$) were tested for the selection.
All columns except for \texttt{i\_mask\_s18a\_bright\_objectcenter} are
from the \texttt{s18\_wide.forced} table. In addition,
\texttt{i\_mask\_s18a\_bright\_objectcenter}
is from the \texttt{s18\_wide.masks} table.
\end{tabnote}
\end{table}

The S18a dataset also provides 
photo-$z$
and the stellar mass for
most of the sources as an ancillary catalog~\citep{Tanaka+18}.
We utilized the 
photo-$z$
and stellar masses calculated using
the Direct Empirical Photometry code \citep[DEmp:][]{Hsieh+14}.
The photo-$z$ and stellar mass were computed from HSC photometry
using the empirical fitting method independently.

\begin{figure}
  \begin{center}
    \includegraphics[width=0.53\textwidth]{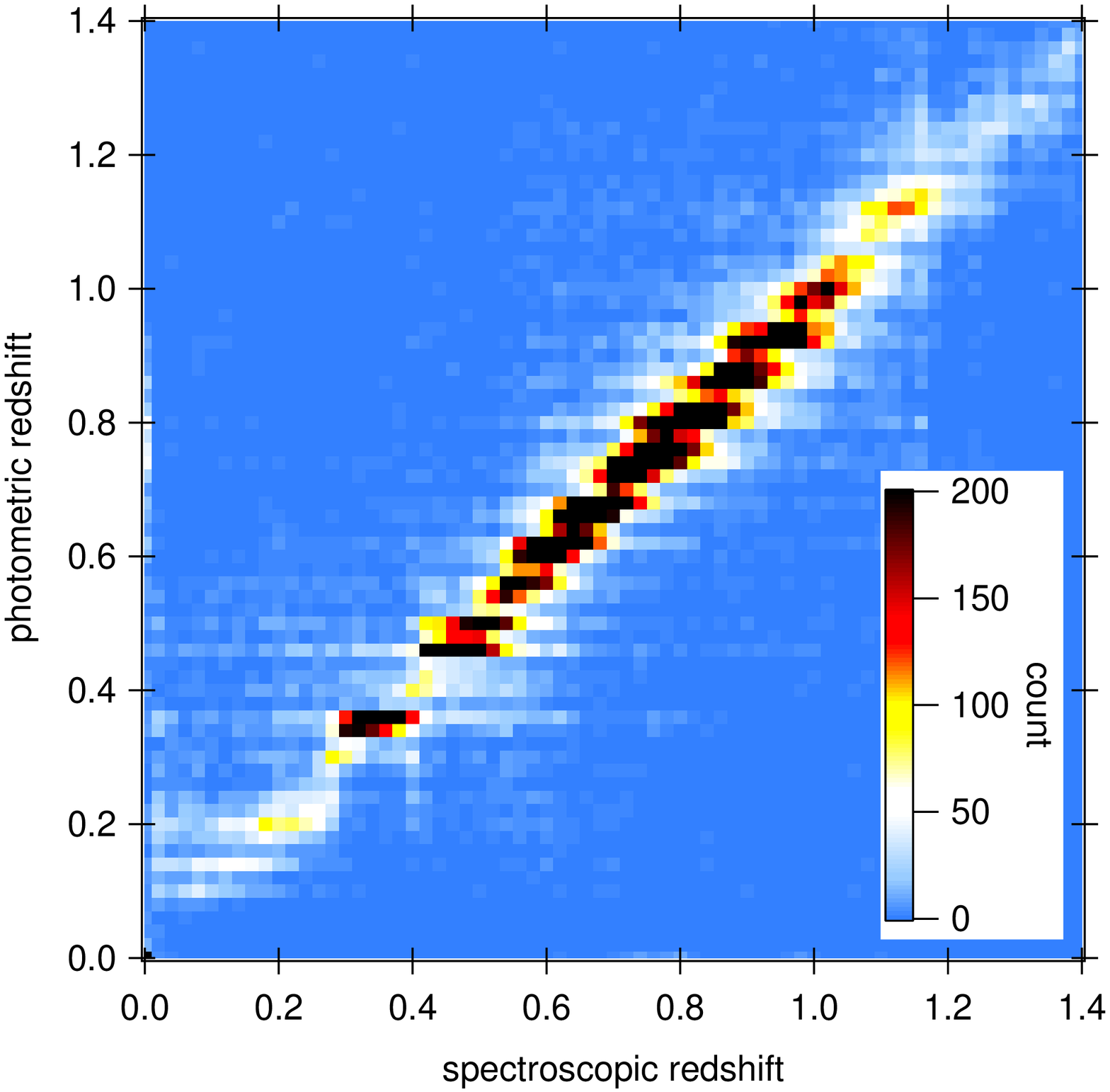}
    \includegraphics[width=0.40\textwidth]{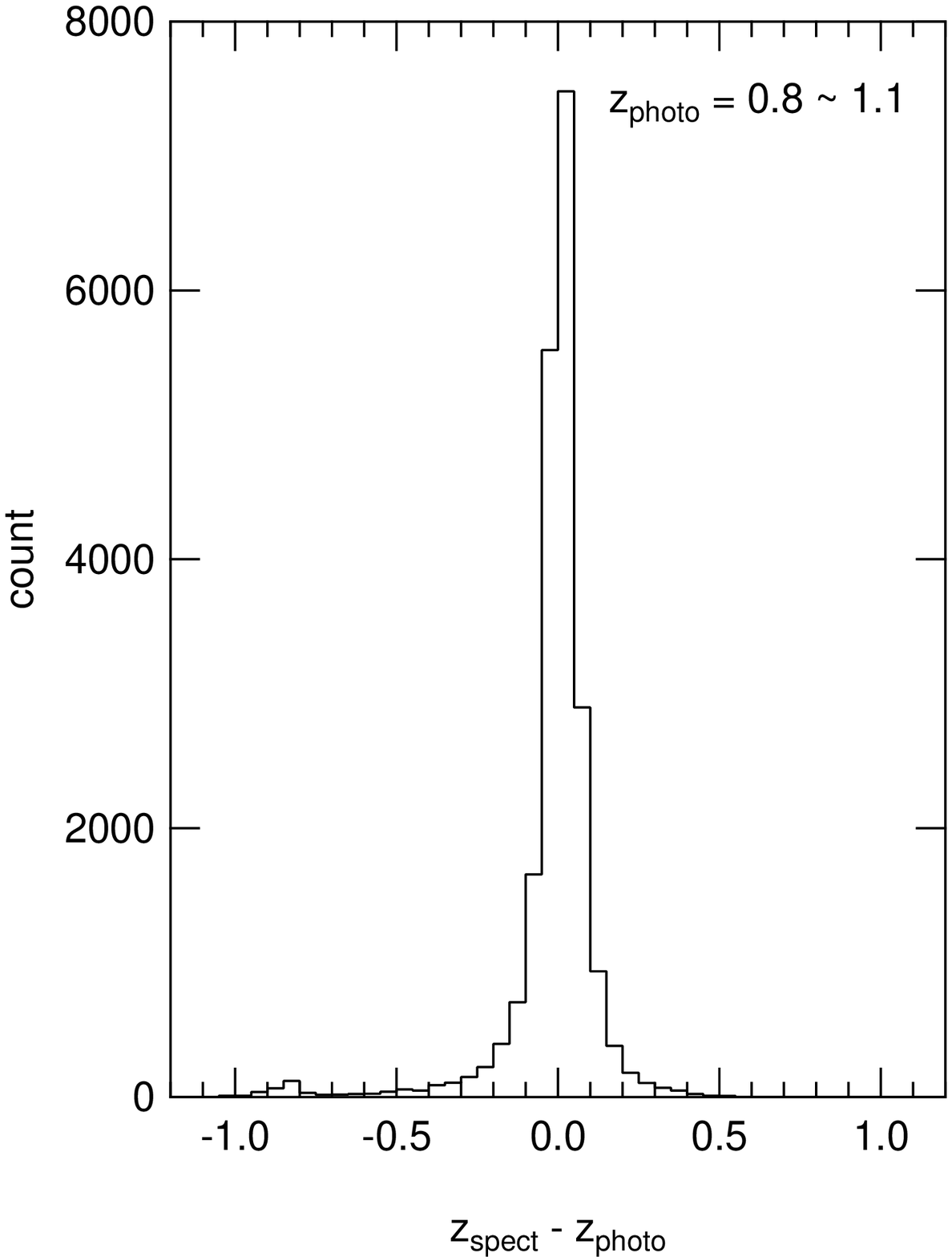}
  \end{center}
  \caption{Comparison of photo-$z$ derived from HSC photometric 
data and spec-$z$  drawn from redshift catalogs of SDSS DR14, 
WiggleZ, DEEP2 DR4, VVDS, VIPERS, and PRIMUS. The left panel shows a density plot
of photo-$z$ vs spec-$z$. 
The right panel shows a histogram
of difference between the two redshifts for objects at photo-$z$
of 0.8--1.1.
}
  \label{fig:comp_z}
\end{figure}

In figure~\ref{fig:comp_z} we compared the photo-$z$ ($z_{\rm photo}$) 
with the spectroscopic redshifts (spec-$z$, $z_{\rm spect}$) drawn from 
SDSS DR14~\citep{Abolfathi+18}, 
WiggleZ final~\citep{Drinkwater+18}, 
DEEP2 DR4~\citep{Newman+13}, 
VVDS~\citep{Fevre+13}, 
VIPERS~\citep{Scodeggio+18}, 
and PRIMUS~\citep{Coil+11,Cool+13}
The matching of the objects were performed by searching nearest neighbors
within 1 arcsec from the HSC sources.
The comparison were made for HSC sources with $i$-band magnitudes between 22 and 
24~mag with average magnitude of 22.6~mag.
This magnitude range was chosen to match with a typical brightness range in
this work.
The standard deviation of the differences is 0.084 after three sigma clipping
for objects of $z_{\rm photo} = $ 0.8 -- 1.1, and the fraction
of outlier is 18\% if it is defined as the fraction of
$|z_{\rm spect} - z_{\rm photo}| > 0.1$.

\begin{figure}
  \begin{center}
    \includegraphics[width=0.6\textwidth]{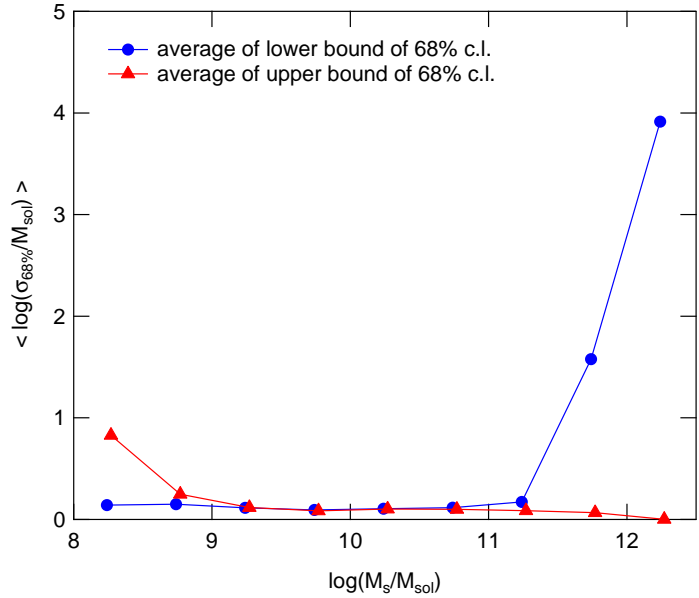}
  \end{center}
  \caption{
Averages of errors in the estimates of stellar mass
for lower (solid circles connected with lines) and 
upper (solid triangles connected with lines) bound of 68\% C.L. 
interval
in $\log{M_{s}/M_{\odot}}$.
}
  \label{fig:stmass-err}
\end{figure}

The averages of errors in the estimates of stellar mass are plotted in 
figure~\ref{fig:stmass-err} for lower and upper bound of 68\% confidence 
interval.
The error of stellar mass is 0.1 dex at $M_{\rm s} = $ 10$^{9}$ -- 10$^{11}$M$_{\odot}$, 
whereas the error of lower bound rapidly increases above 10$^{11.5}$M$_{\odot}$.

\subsection{AGN samples}
\label{sec:sample_agn}

The AGNs were drawn from the
Million Quasars (MILLIQUAS) catalog v5.7 2019 update \citep{Flesch-15}.
MILLIQUAS is a compilation of identified AGNs/QSOs or their candidates
from various studies and QSO catalogs, which have reached
1 983 749 in number.
We selected AGNs for which 
spec-$z$ are within the range 
of 0.8--1.1 and that are well contained within the area of
the S18a HSC-SSP wide dataset.
The selected AGNs were further filtered according to the conditions of 
the HSC sources around them and their proximity, as described in 
section~\ref{sec:dataset_selection}.

\begin{figure}
  \begin{center}
    \includegraphics[width=0.6\textwidth]{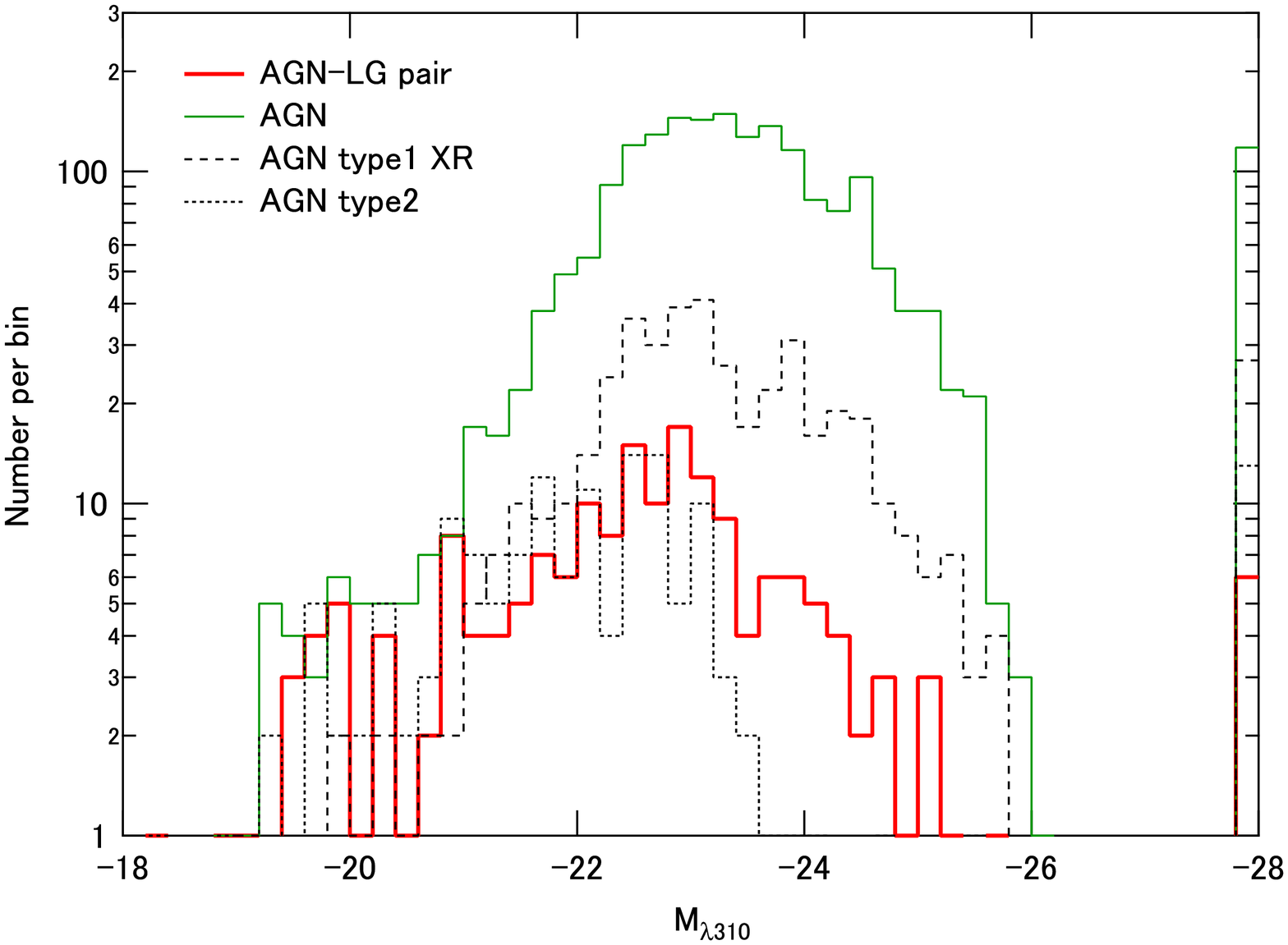}
  \end{center}
  \caption{Distribution of absolute magnitude $M_{\lambda 310}$ for each AGN sample.
AGNs for which the absolute magnitude was not measured owing to saturation or other
selection criteria are counted at the brightest bin. }
  \label{fig:hist_QSO_M310}
\end{figure}

We extracted four AGN samples, namely,
AGNs from all types (hereinafter referred to as
a whole AGN sample or simply an AGN sample), 
AGNs with an associated nearby luminous galaxy
($M_{\lambda 310} < -21$)
(an AGN-LG pair sample or simply an AGN-LG sample),
type~1 AGNs through X-ray or radio emissions 
(an AGN type~1 XR sample),
and type~2 AGNs (an AGN type~2 sample).
The absolute magnitude $M_{\lambda 310}$ distributions for these AGN samples 
are shown in figure~\ref{fig:hist_QSO_M310}.
The method for calculating the absolute magnitude is the same as
that described in \citet{Shirasaki+18}, and is detailed in the following section
as well.
For some of the AGNs, we were unable to measure $M_{\lambda 310}$ owing
to a saturation in the HSC photometry or other selection criteria preventing
the source to be analyzed.
Such AGNs are counted at the brightest bin at $M_{\lambda 310} = -28.0$ -- $-27.8$.

\begin{figure}
  \begin{center}
    \includegraphics[width=0.6\textwidth]{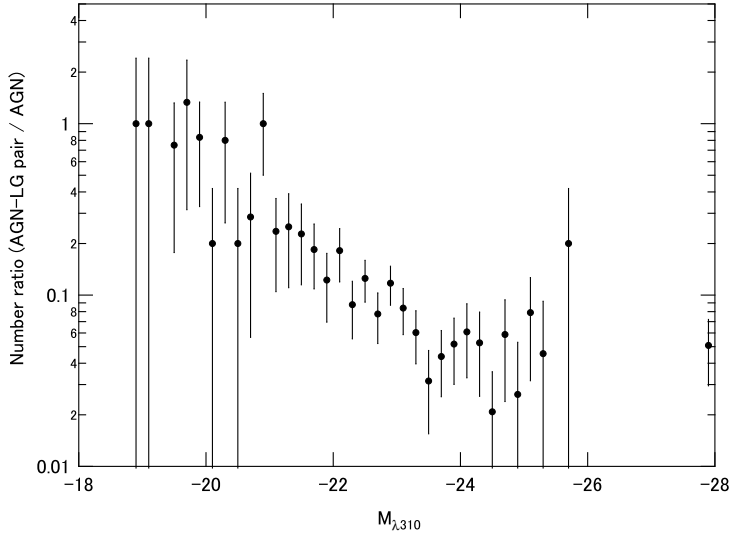}
  \end{center}
  \caption{Ratio of the number of
    AGNs in the AGN-LG sample
    to their number in the whole AGN sample
    at each absolute magnitude $M_{\lambda 310}$.}
  \label{fig:hist_ratio_M310}
\end{figure}

To see the difference in the distributions of the absolute magnitude between
the AGN-LG sample and whole AGN sample,
their number ratios are plotted
in figure~\ref{fig:hist_ratio_M310}.
The result shows that the AGNs in the AGN-LG sample are dominated by lower
luminosity AGNs as compared to those in the 
whole AGN sample.

In constructing the sample of AGN-LG pairs, LGs were 
drawn from the following six redshift survey catalogs:
SDSS DR14~\citep{Abolfathi+18}, 
WiggleZ final~\citep{Drinkwater+18}, 
DEEP2 DR4~\citep{Newman+13}, 
VVDS~\citep{Fevre+13}, 
VIPERS~\citep{Scodeggio+18}, 
and PRIMUS~\citep{Coil+11,Cool+13}.
These catalogs were also used for 
constructing the blue galaxy sample.

\begin{figure}
  \begin{center}
    \includegraphics[width=0.6\textwidth]{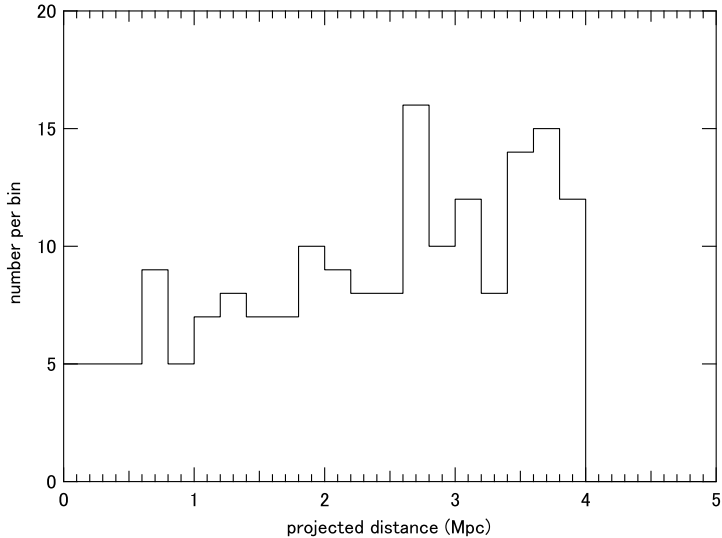}
  \end{center}
  \caption{Distribution of projected separation distance between AGN
    and an LG in the AGN-LG pair sample.}
  \label{fig:hist_drp}
\end{figure}

The absolute magnitude $M_{\lambda 310}$ in these catalogs was measured for each galaxy,
as described in the following section.
The counterpart LGs were searched within 4~Mpc in projected distance from the AGN 
and 5~Mpc in the line-of-sight direction determined from the redshift measurement.
The line of sight distance was set 1~Mpc larger to accommodate the
redshift uncertainty.
The typical uncertainties of redshift are 0.00006 (DEEP2), 0.0003 (SDSS, WiggleZ),
0.001 (VIPERS), 0.0014 (VVDS), and 0.01 (PRIMUS).
Considering that 50\% of the the AGN-LG pairs come from the DEEP2, SDSS, or
WiggleZ catalog, the margin of distance was chosen to be 1~Mpc, which corresponds
to the redshift interval of 0.0004.
The distribution of the separation distances to the brightest LGs is shown
in figure~\ref{fig:hist_drp}.

To reduce the effect of the redshift dependence of the clustering and other
environmental properties in the comparison among the different samples, 
a redshift-match-selection was carried out by selecting the same relative 
number of target objects at random for each redshift range of $z=$0.8--0.9, 0.9--1.0,
and 1.0--1.1.
The relative numbers were determined from the numbers in the AGN-LG sample, 
which are 76, 54, and 50 for those redshift ranges, respectively.
The details of the sample selections other than those described herein are provided in section~\ref{sec:dataset_selection}.

The numbers of AGNs for each sample are summarized in 
table~\ref{tab:stat_agn} for each AGN type.
As was previously shown in figure~\ref{fig:hist_ratio_M310}, AGNs in the
AGN-LG sample are biased toward lower luminosity AGNs, classified
as ``AGN'' rather than ``QSO'', when compared to those in the
whole AGN sample.
  The classification between AGN and QSO in the MILLIQUAS catalog is
  in principle based on their morphology~\citep{Flesch-15};
  The core-dominated objects with no disk seen are classified as QSO
  and disk dominated ones are classified as AGN.
  For faint and unresolved objects, the classification is made by
  their apparent magnitude; The objects brighter than the following
  magnitude are classified as QSO~\citep{Flesch-15}:
  \begin{equation}
    m = 22 + 5 \log_{10}{(z (0.8 + z) - 0.09)},
  \end{equation}
  where $z$ is a redshift of the object.
The QSO-to-AGN ratio for the AGN-LG sample is 3.9:1, whereas
that for the whole AGN sample is 25.8:1.

\begin{table}
\tbl{Summary of the number of each AGN type for four AGN samples.}{
\begin{tabular}{ccccrrrr}
\hline
\multicolumn{3}{c}{type}      & label$^{a}$    & $n_{\textrm{AGN}}^{b}$
                                            & $n_{\textrm{AGN-LG}}^{c}$ 
                                            & $n_{\textrm{Type1-XR}}^{d}$ 
                                            & $n_{\textrm{Type2}}^{e}$  \\ 
\hline
AGN & type I   &             & A            & 21  & 11 &  0 &  0 \\
    &          & X-ray       & AX           & 12  &  4 & 16 &  0 \\
    & type II  &             & N            & 11  & 10 &  0 & 17 \\
    &          & X-ray       & NX           & 27  & 12 &  0 & 36 \\
    &          & Radio       & NR           &  1  &  0 &  0 &  1 \\
    &          & X \& Radio  & NRX          &  1  &  0 &  0 &  1 \\
QSO & type I   &             & Q            &1420 & 37 &  0 &  0 \\
    &          & X-ray       & QX           & 314 & 93 &363 &  0 \\
    &          & Radio       & QR,QR2       &  53 &  0 & 51 &  0 \\
    &          & X \& Radio  & Q2X,QRX,QR2X &  20 &  1 & 20 &  0 \\
    & type II  &             & K            & 59  &  7 &  0 & 70 \\
    &          & X-ray       & KX           & 16  &  4 &  0 & 23 \\
    &          & Radio       & KR           &  1  &  0 &  0 &  1 \\
    &          & X \& Radio  & KRX          &  0  &  1 &  0 &  1 \\
\hline
Total &        &             &              &1956 &180 &450 &150 \\
\hline
\end{tabular}}\label{tab:stat_agn}
\begin{tabnote}
  $^{a}$ classification of object in the MILLIQUAS catalog:
  Q = QSO type-I,
  A = AGN type-I,
  K = narrow line QSO type-II,
  N = narrow line AGN type-II,
  R = radio association,
  X = X-ray association,
  2 = double radio lobes.
  $^{b}$ number in the whole AGN sample.
  $^{c}$ number in the AGN-LG pair sample.
  $^{d}$ number in the AGN type~1 XR (X-ray and/or radio detection) sample.
  $^{e}$ number in the AGN type~2 sample.
\end{tabnote}
\end{table}
\subsection{Blue $M_{*}$ galaxies}
\label{sec:sample_d210}

Blue $M_{*}$ galaxies 
were drawn from the following six redshift 
survey catalogs:
SDSS DR14, WiggleZ final, DEEP2 DR4, VVDS, VIPERS, and PRIMUS.
We selected galaxies located within the HSC-SSP wide area and
within the redshift range of 0.8--1.1.
This sample is used as a representative of ordinary galaxies located
at a relatively smaller density field than those of the AGNs and red galaxies.

We identified the HSC sources corresponding to those galaxies, and
measured their color and absolute magnitude using the HSC photometric 
data.
The measurement was made by fitting the galaxy SED templates to the observed
SEDs using EAZY software developed by \citet{Brammer+08}.
The absolute magnitude $M_{\lambda 310}$ was measured at the rest frame 
wavelength of 310~nm, and the color was defined as $D = M_{\lambda 270} - M_{\lambda 380}$,
where $M_{\lambda 270}$ and $M_{\lambda 380}$ are the absolute magnitude
at 270 and 380~nm, respectively.

We then selected blue $M_{*}$ galaxies that satisfy the criteria 
$M_{\lambda 310} \ge -21$ and $D < 1.4$.
As shown in the later results section, the color distribution for galaxies
is well represented with a linear combination of three Gaussian functions,
which represent blue clouds, red sequences, and green valley galaxies.
The criterion $D < 1.4$ separates most of the red galaxies
from the sample.

In addition to the selections based on the properties of the target itself,
the coverage and uniformity of the HSC sources around the targets
were taken into account.
As conducted for the AGN samples, a redshift-match-selection was applied.
The details of the sample selection other than those explained herein
will be described in section~\ref{sec:dataset_selection}.
In total, 2 426 blue $M_{*}$ galaxies are used in this analysis.
The number of galaxies from each survey is summarized in table~\ref{tab:stat_d210}.

\begin{table}
\tbl{Summary of the number of galaxies from each survey for the blue galaxy sample .}{
\begin{tabular}{ccccccc}
\hline
SDSS  & WiggleZ & DEEP2 & VVDS & VIPERS & PRIMUS & total \\ 
\hline
106   &    56   &   430 & 123  & 946    &  765   & 2426 \\
\hline
\end{tabular}}\label{tab:stat_d210}
\end{table}

\subsection{Dataset selection}
\label{sec:dataset_selection}

In the analysis described herein, we treated each target object (AGN or blue galaxy) 
and its surrounding HSC sources as a set.
Hereafter, we refer to the unit of the dataset simply as a dataset.
In this section, we describe the criteria to include in the datasets
for analysis.
To homogenize the environments of the target objects as much as 
possible and avoid an edge effect of the survey boundary, we
selected target objects that are well within the survey footprint.
To do so, we measured the radial distribution 
of random sources from the position of the target objects.
The random source catalog is available for every data release.
The random sources were generated with a density of 100 per square arcmin
inside the survey footprint, allowing us to estimate the fraction of unobserved
 or masked regions by counting the random sources.

We selected the random sources adapting the same criteria adapted 
to the real data if applicable.
Their radial distribution was measured in annuli spaced by 0.2~Mpc out to 
10~Mpc from the targets.
We kept only those targets around which $>$60\% of the area of all annuli at 
$\ge$2~Mpc and $>$80\% of the area at $<$2Mpc were included in the survey 
footprint and not masked for bright sources.
The procedure to build and validate the bright-star masks for the 
HSC-SSP survey is described in~\citet{Coupon+18}.
The datasets that passed this selection numbered
2 740 for the AGNs and 5 222 for the blue galaxies.

The spatial uniformity of the HSC sources around the targets was also
examined to identify the datasets that are significantly affected by 
a high-density foreground region, spurious sources around bright stars,
and so on.
For this purpose, we calculated two parameters for the radial 
number density distribution of the galaxies $\chi^{2}$ and
$\sigma_{\rm max}$, where $\chi^{2}$ is a square sum
of the deviation from the number density distribution fitted to the
observed data using equation~(\ref{eq:density_r}), which is derived
in section~\ref{sec:cc_galaxy},
where $\sigma_{\rm max}$ is the maximum deviation from the density distribution.
The adapted criteria for these parameters are as follows:
$\chi^{2}/n \le 4$ and $\sigma_{\rm max} \le 6$.
These criteria was chosen to remove a few \% of the datasets that deviate
the most from a uniform distribution.
  We checked the effect of these criteria to the estimate of
  cross-correlation length and confirmed that difference
  between the cross-correlation lengths calculated for the datasets
  for which these selection are adapted or not is within the
  statistical error.
The datasets that passed all selections described above 
totaled 2 720 for the AGNs and 5 126 for the blue galaxies.

In the analysis adapted for this study, it is crucial to construct the 
datasets such that the contribution from the foreground and background galaxies 
are smeared out by stacking the radial number distribution of the galaxies.
To avoid stacking numerous identical fields, we selected
the target objects 
so that they
have no more than one other target object
within 4~Mpc.
This selection was adapted for each redshift bin, which was divided into $z = $ 
0.8--0.9, 0.9--1.0, and 1.0--1.1.
The datasets that passed all selections described above 
numbered 2 622 for the AGNs and 4 262 for the blue galaxies.

To reduce the effect of the redshift dependence in the
comparison of the environmental properties, we selected the dataset
such that the relative number of datasets for three redshift bins 
becomes the same among the five target groups.
The final numbers of datasets that passed all selections are
summarized in table~\ref{tab:stat_final}
along with the median
redshifts and absolute magnitudes of the datasets.
The numbers of the datasets broken down by the HSC survey fields are
also summarized in table~\ref{tab:stat_by_field}.

\begin{table}
    \tbl{
    Summary of the number, median redshift, and median absolute magnitude
    in the sample for each target type and each redshift group.
    }{
\begin{tabular}{ccrcc}
\hline
$z^{a}$  & sample type
         & $n^{b}$
         & $\tilde{z}^{c}$ 
         & $\tilde{M}_{\lambda 310}^{d}$  \\
\hline
0.8--0.9 & AGN           &  826 &  0.85 & $-$23.0 \\
         & blue galaxy   & 1024 &  0.85 & $-$20.6 \\
         & AGN-LG pair   &   76 &  0.85 & $-$22.3 \\
         & AGN type~1 XR &  190 &  0.84 & $-$22.8 \\
         & AGN type~2    &   63 &  0.85 & $-$21.7 \\  
\hline
0.9--1.0 & AGN           &  587 &  0.95 & $-$23.4 \\
         & blue galaxy   &  728 &  0.93 & $-$20.7 \\
         & AGN-LG pair   &   54 &  0.96 & $-$22.9 \\
         & AGN type~1 XR &  135 &  0.96 & $-$23.2 \\
         & AGN type~2    &   45 &  0.96 & $-$22.4 \\
\hline
1.0--1.1 & AGN           &  543 &  1.05 & $-$23.8 \\
         & blue galaxy   &  674 &  1.04 & $-$20.7 \\
         & AGN-LG pair   &   50 &  1.03 & $-$22.9 \\
         & AGN type~1 XR &  125 &  1.05 & $-$23.8 \\
         & AGN type~2    &   42 &  1.04 & $-$22.6 \\
\hline
0.8--1.1 & AGN           & 1956 &  0.92 & $-$23.3 \\
         & blue galaxy   & 2426 &  0.92 & $-$20.7 \\
         & AGN-LG pair   &  180 &  0.91 & $-$22.6 \\
         & AGN type~1 XR &  450 &  0.93 & $-$23.2 \\
         & AGN type~2    &  150 &  0.94 & $-$22.1 \\
\hline
      \end{tabular}}
  \label{tab:stat_final}
\begin{tabnote}
  $^{a}$~redshift range.
  $^{b}$~number of datasets.
  $^{c}$~median redshift.
  $^{d}$~median absolute magnitude $M_{\lambda 310}$.
  The median redshift and absolute magnitude for AGN-LG pair
  are calculated for the AGN.
\end{tabnote}
\end{table}

\begin{table}
    \tbl{
    Summary of the number of the datasets for each target type and
    HSC survey field.
    }{
\begin{tabular}{crrrrr}
\hline
field name & $n_{\rm AGN}^{a}$  & $n_{\rm G}^{b}$
           & $n_{\rm AGN-LG}^{c}$ & $n_{\rm Type1-XR}^{d}$
           & $n_{\rm Type2}^{e}$ \\
\hline
WIDE12H/GAMA15H & 583 &  36 &   3 &  70 & 31 \\
VVDS            & 490 & 626 &  38 &  58 & 13 \\
GAMA09H         & 259 & 252 &  18 &  42 & 61 \\
XMM-LSS         & 376 &1358 & 102 & 211 & 44 \\
HECTOMAP        & 166 &   0 &   3 &  26 &  0 \\
WIDE01H         &  67 &   1 &   2 &  28 &  0 \\
AEGIS           &  15 & 153 &  14 &  15 &  1 \\
\hline
      \end{tabular}}
  \label{tab:stat_by_field}
\begin{tabnote}
  $^{a}$~number in the whole AGN (for all types) sample.
  $^{b}$~number in the blue galaxy sample.
  $^{c}$~number in the AGN-LG pair sample.
  $^{d}$~number in the AGN type~1 XR (X-ray and/or radio detection) sample.
  $^{e}$~number in the AGN type~2 sample.
\end{tabnote}
\end{table}

\section{Analysis method}\label{sec:analysis}

\subsection{Cross-correlation between targets and HSC sources.}
\label{sec:cc_galaxy}

The cross-correlation functions between the target objects and HSC sources
were calculated using the method described in our previous
papers~\citep{Shirasaki+11,Komiya+13,Shirasaki+16,Shirasaki+18}, which is briefly described herein.

When the redshifts of the target objects are known, we can calculate
the number densities of the HSC sources as a function of the projected distance 
from the target in its redshift plane.
Thanks to the clustering properties of galaxies, the galaxies located
at the target's redshift emerge as an excess over the flat distribution
of the foreground/background galaxies after stacking the radial number 
densities for many of the targets.

The cross-correlation function $\xi(r)$ is a measure of the clustering as an 
excess over random distribution, and is related to the 
number density $\rho(r)$ of the correlated objects (HSC sources
in this analysis) around the target objects (AGN or blue galaxy) as follows:
\begin{equation}
\xi(r) = \rho(r) / \rho_{0} - 1,
\end{equation}
where $\rho_{0}$ is the average number density of the correlated objects
at the redshift of the targets.

Owing to a lack of precise measurements of the distances to the HSC sources along the line of
sight, the projected correlation function $\omega(r_{p})$ is measured instead of
$\xi(r)$ as follows:
\begin{equation}
\omega(r_{p}) = 
\int_{-\infty}^{\infty} \xi(r_{p},\pi) d\pi \simeq
   \frac{n(r_{p}) - n_{bg}}{\rho_{0}},
   \label{eq:omega_0}
\end{equation}
where 
$r_{p}$ and $\pi$ are distance from a target object
perpendicular and along to the line of sight, respectively, and
$n(r_{p})$
$ = \int_{-\infty}^{\infty} \rho(r_{p},\pi) d\pi$
represents the average surface number density
of the HSC sources at a projected distance $r_{p}$, and $n_{\mathrm{bg}}$ 
$ = \int_{-\infty}^{\infty} \rho_{0}(\pi)d\pi$
represents the average density expected for the case in which all
galaxies are uniformly distributed.
In deriving the right-hand side expression of equation~(\ref{eq:omega_0}), we used
approximation that the effective integral interval is limited to $\pi \sim 0$.

According to the measurements of the galaxy auto-correlation function described in the literature, 
the correlation function is approximated using a power law function, i.e., 
$\xi(r) = (r_{0}/r)^{\gamma}$.
The typical value of the power index $\gamma$ is 1.8~\citep[e.g.][]{Zehavi+11,Coil+08,Coil+17}, 
and $r_{0}$ is termed the correlation length, which is a measure of the galaxy clustering.
In this case, $\omega(r_{p})$ is expressed as follows:
\begin{equation}
   \omega(r_{p}) 
      = r_{p} \left( \frac{r_{0}}{r_{p}} \right)^{\gamma} 
      \frac{\Gamma(\frac{1}{2})\Gamma(\frac{\gamma-1}{2})}{\Gamma(\frac{\gamma}{2})}.
   \label{eq:omega_1}
\end{equation}
Equating the right-hand sides of equations (\ref{eq:omega_0}) and (\ref{eq:omega_1}), 
$n(r_{p})$ is expressed as follows:
\begin{equation}
   n(r_{p}) = r_{p} \left( \frac{r_{0}}{r_{p}} \right)^{\gamma} 
      \frac{\Gamma(\frac{1}{2})\Gamma(\frac{\gamma-1}{2})}{\Gamma(\frac{\gamma}{2})}
      \rho_{0} + n_{\mathrm{bg}}.
   \label{eq:density_r}
\end{equation}
$\rho_{0}$
for each dataset is
calculated 
from the luminosity function, which was derived by parametrizing
the luminosity functions in the literature, and the completeness function
$C(m)$.
The detail of the parametrization of the luminosity function
and the completeness function is
described in \citet{Shirasaki+18}
(completeness function is refereed to as detection efficiency $DE(m)$ in the reference).
The average 
for all the datasets in the sample
provides $\rho_{0}$ in equation~(\ref{eq:density_r}).

The 
completeness as a function of magnitude
$C(m)$
is required to correct for the completeness by multiplying $C(m)$ to the 
luminosity function model in deriving $\rho_{0}$.
It
was calculated as a ratio of the observed 
magnitude distribution $N_{\mathrm{obs}}(m)$ to the model function $N_{\mathrm{org}}(m)$, 
which is a magnitude distribution expected for an ideal observation of a 100\% 
completeness at any magnitude.
For $N_{\mathrm{org}}(m)$, we assumed a broken power law form, and the power law 
index was
determined using data from the HSC-SSP S18a deep survey dataset of the COSMOS field.
Next, 
$C(m)$
was determined for each dataset by fitting 
$C(m)$$N_{\mathrm{org}}(m)$
to $N_{\mathrm{obs}}(m)$.
For a model of 
$C(m)$, the same functional form as defined in equation~(14)
of \citet{Shirasaki+18} was used.

\begin{figure}
  \begin{center}
    \includegraphics[width=0.6\textwidth]{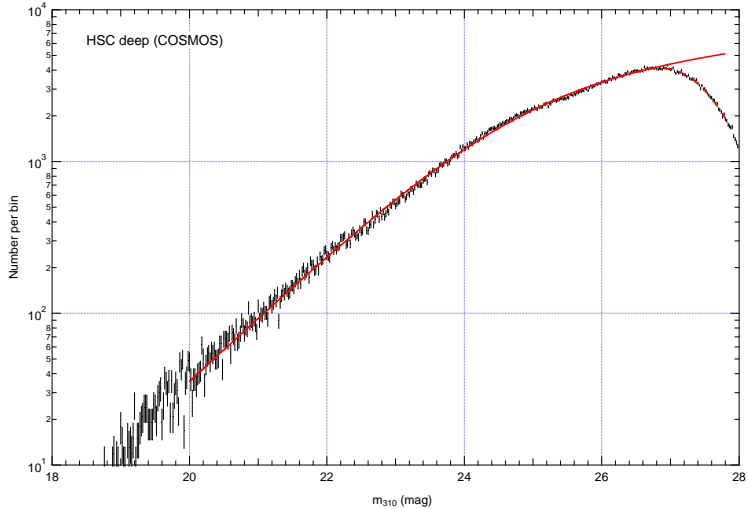}
  \end{center}
  \caption{
Distribution of apparent magnitude $m_{\lambda 310}$ derived from
a deep dataset of the COSMOS field, where $m_{\lambda 310}$ is an apparent 
magnitude measured at wavelength 310$(1+0.9)$~nm in the observer frame, that is
310~nm in the rest frame at redshift 0.9. 
The solid and dashed lines represent the fitted function expressed in a broken 
power law function and that multiplied by completeness
$C(m)$, respectively.
}
  \label{fig:hist_mag_dud}
\end{figure}

The accuracy of the model function is demonstrated in figure~\ref{fig:hist_mag_dud}
for the magnitude distribution derived from a deep dataset of the COSMOS field.
The observed magnitude distribution is well fitted with the broken power law model 
at magnitudes $m_{\lambda 310} < 26.8$~mag.
Since this work is performed at magnitudes $m_{\lambda 310} < 26$~mag for the wide 
dataset, it is reasonable to assume a broken power law form for $N_{\mathrm{org}}(m)$.

The radial distributions $n(r_{p})$ of the HSC sources were derived 
by stacking the distributions for all the datasets in the sample,
where HSC sources with absolute magnitudes of $M_{\lambda 310} < -19$~mag were used.
This threshold magnitude corresponds to an approximately 90\% 
detection (completeness) limit.
We applied this analysis for the photo-$z$ selected galaxies, 
which were constructed by selecting the HSC source whose photo-$z$
are within the range of $\pm 0.1$ from the redshifts of the target object.
Although the completeness is reduced by this photo-$z$ selection, 
background/foreground galaxies are also reduced more efficiently, which results
in an increase in the signal to noise ratio of the clustering.

\begin{figure}
  \begin{center}
    \includegraphics[width=0.48\textwidth]{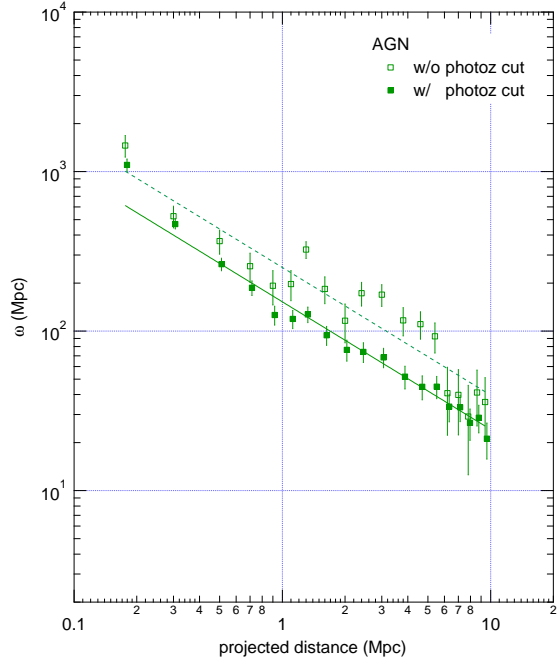}
  \end{center}
  \caption{
    Projected cross-correlation functions for the whole AGN sample.
    The open squares indicate the correlation function obtained using all of the
    galaxies, whereas the solid squares represent that obtained using the photo-$z$ selected
    galaxies. The same $\rho_{0}$ parameter, which was estimated without
    correcting the reduction through a photo-$z$ selection, is used for both. We use
    the ratio between them as the correction factor for the photo-$z$ selection.
  }
  \label{fig:cc_AGN}
\end{figure}

The reduction in the completeness 
by the photo-$z$ 
selection was
estimated by comparing the projected cross-correlation functions calculated
for galaxies with and without photo-$z$ selection.
The two projected cross-correlation functions were calculated for the same
$\rho_{0}$ parameter, which was an estimate for the galaxies without a photo-$z$
selection, and the average ratio between them was then taken to be the
reduction rate owing to the photo-$z$ selection.
The reduction rate was estimated to be 0.61 for the whole
AGN sample, which has
large statistical sample size and provides the best signal-to-noise ratio for the
clustering, and was also used for the other samples.
In figure~\ref{fig:cc_AGN}, the two projected cross-correlation functions
used to derive the reduction rate are shown.

The
effective observational area, 
which is used for the calculation of $n(r_{p})$ and
is an area corrected for a dead region affected by a bright source,
survey boundary, gaps between observations, or other factors,
was estimated using a random catalog.
The random catalog, 
which was created at random positions avoiding the dead region with 
number density 100 arcmin$^{-1}$,
was extracted from the S18a database.
We fixed $\gamma$ to 1.8, and the model function given by the equation 
(\ref{eq:density_r}) is fitted to the observed radial surface densities
$n(r_{p})$ with two free parameters $r_{0}$ and $n_{\mathrm{bg}}$.
The fitting was performed by the least square method by weighting
each data point with a inverse square of the error determined by Poisson 
statistics.
Since the target objects were selected so as to avoid overlap of the environment
regions among them as much as possible, each bin of
  $n(r_{p})$
is almost independent
from each other and the covariance between them is negligible.
Thus we ignored the covariance in the fitting.
The cross-correlation length is estimated 
as an optimal solution for the free parameter $r_{0}$.
The quoted error is a confidence interval in one sigma unless otherwise stated.

We also cross-correlate the target objects to the cluster of the HSC sources.
Because no reliable model is available for the cluster mass function, we
simply derive the radial number density of the clusters.

\subsection{Color, absolute magnitude, and stellar mass distributions around the targets}
\label{sec:dist_X}

Although we do not have 
spec-$z$ of individual HSC 
sources, we can statistically
estimate the distribution of a property X of the HSC sources located 
at distance within a few Mpc from the target object.
Thanks to the clustering feature of galaxies, their number density
increases as we get closer to the target objects.
Thus, subtracting the distribution of property X measured in a lower density region
from that measured at a higher density region, we can estimate the net distribution
of property X for HSC sources associated with the target objects.

Property X can be anything that is measurable for the HSC sources.
In this analysis, we investigate the distribution of color, absolute magnitude,
and stellar mass of the HSC sources.
The color is calculated as the difference between magnitudes at the rest frame
wavelengths of 270 and 380~nm, and the absolute magnitude is measured at 
310~nm in the rest frame of the target object.
As previously described in sub-section~\ref{sec:sample_d210}, EAZY software \citep{Brammer+08}
was used to interpolate the SED derived from the HSC photometric data.
The stellar mass was obtained from the ancillary catalog available along with
the photo-$z$.
We used those calculated using the DEmp code~\citep{Hsieh+14}.
The photo-$z$ was used to select HSC sources associated
with the target objects.

Although the photo-$z$ selection is useful for increasing the 
signal to noise ratio of the derived distribution, it distorts the intrinsic 
distribution owing to the dependence of the completeness on the examined
property.
Thus the completeness should be corrected taking into account
its dependence on the property, when the distribution needs to be compared 
in an absolute manner, e.g. in a case where it is compared with the luminosity 
functions derived in literature.
The reduction rate by the photo-$z$ selection as a function of
the absolute magnitude is estimated by comparing 
the magnitude distribution obtained for all galaxies (without 
photo-$z$ selection) with that for the photo-$z$ selected galaxies of the
whole
AGN sample.
In addition, we also investigated the peak density distribution of the
clusters detected as stellar mass density peaks (mass peak clusters) and 
clusters detected as number density peaks (number peak clusters), which were found 
using a procedure for detecting peaks in the 2D distribution of the HSC sources.

\subsection{Identification of peak locations of galaxy number density and stellar mass density 
around the targets}
\label{sec:cluster_finding}

As described in section~\ref{sec:introduction}, the strong cross-correlation
between AGNs and LGs found by \citet{Shirasaki+18} extends over 
$\sim$10~Mpc scale, which indicates that the mechanism is related to the
activity in the large-scale structure.
A cluster-cluster interaction is one of the candidates to produce strong
cross-correlation in such a large scale.
Thus we examined the environment of AGN-LG pairs based on the statistics related
to the clusters around them.
The peak locations of the number density and stellar mass density of the HSC 
sources were detected 
to select galaxy cluster candidates
by searching the local maxima in a blurred density map.
The blurred density map was constructed by blurring the positional distribution
of the photo-$z$ selected HSC sources using a 2D Gaussian with $\sigma=$1~Mpc.
The stellar mass density map was created by weighting each HSC source with
its stellar mass, whereas the number density map was created with an equal weight.
The stellar mass was limited to 10$^{12}$~$M_{\solar}$ to avoid dominance
from a single large stellar mass object with a large uncertainty
as shown in figure~\ref{fig:stmass-err}, and thus a stellar
mass exceeding 10$^{12}$~$M_{\solar}$ was 
set to 10$^{12}$~$M_{\solar}$.
We checked that lowering the threshold to $10^{11.5}$~$M_{\solar}$
doesn't affect the result and conclusion.
The local sub-peaks found within a 1~Mpc projected distance from the local maximum
were removed from the sample.

\section{Results}

\subsection{Cross-correlation between target objects and HSC sources}

The cross-correlation between the target objects and the HSC sources were
examined for five target types: whole AGN (of all types), 
blue galaxy, AGN-LG pair, AGN type~1 XR (with detection with 
X-ray and/or radio signals), and AGN type~2.

\begin{figure}
  \begin{center}
    \includegraphics[width=0.48\textwidth]{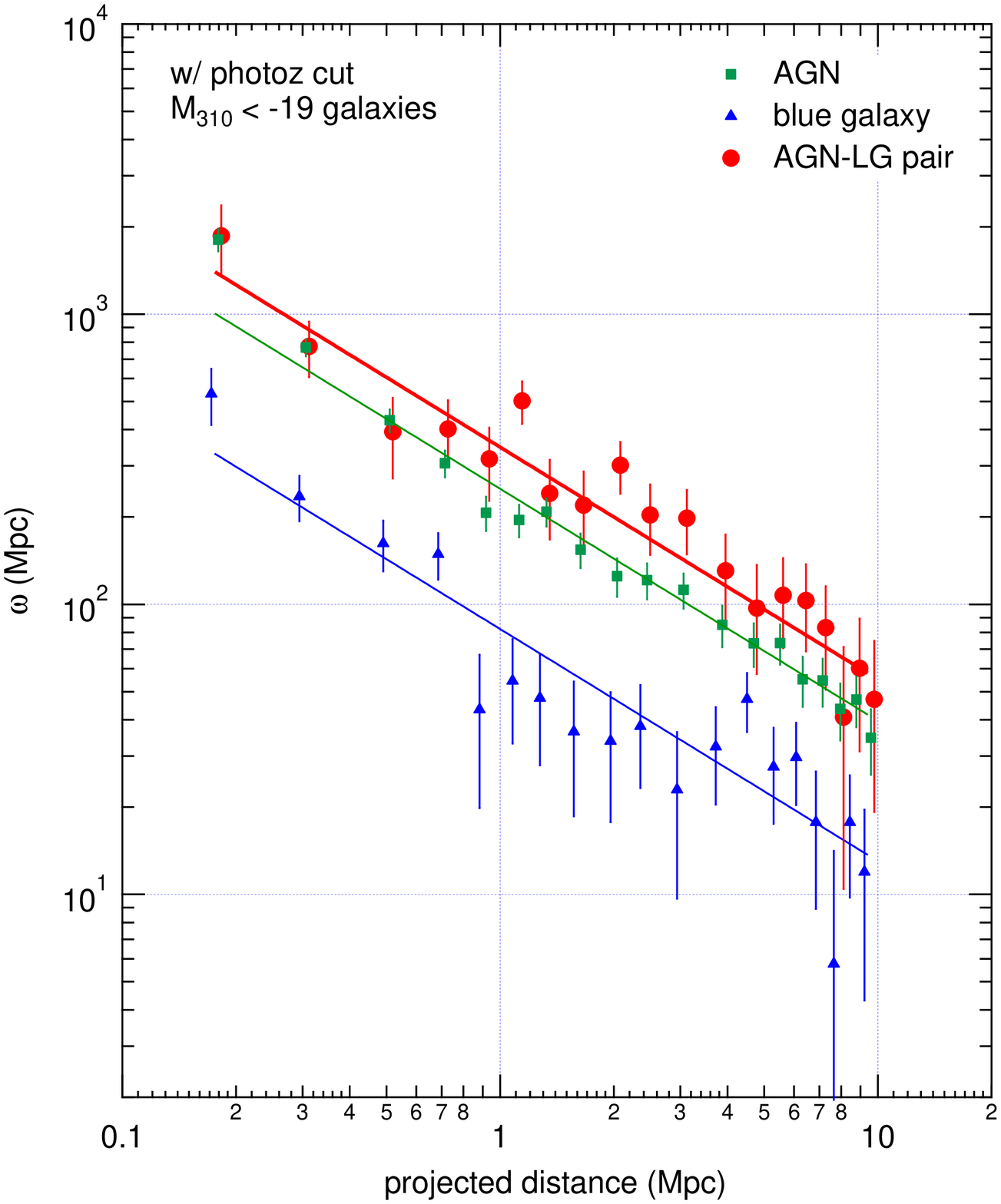}
    \includegraphics[width=0.48\textwidth]{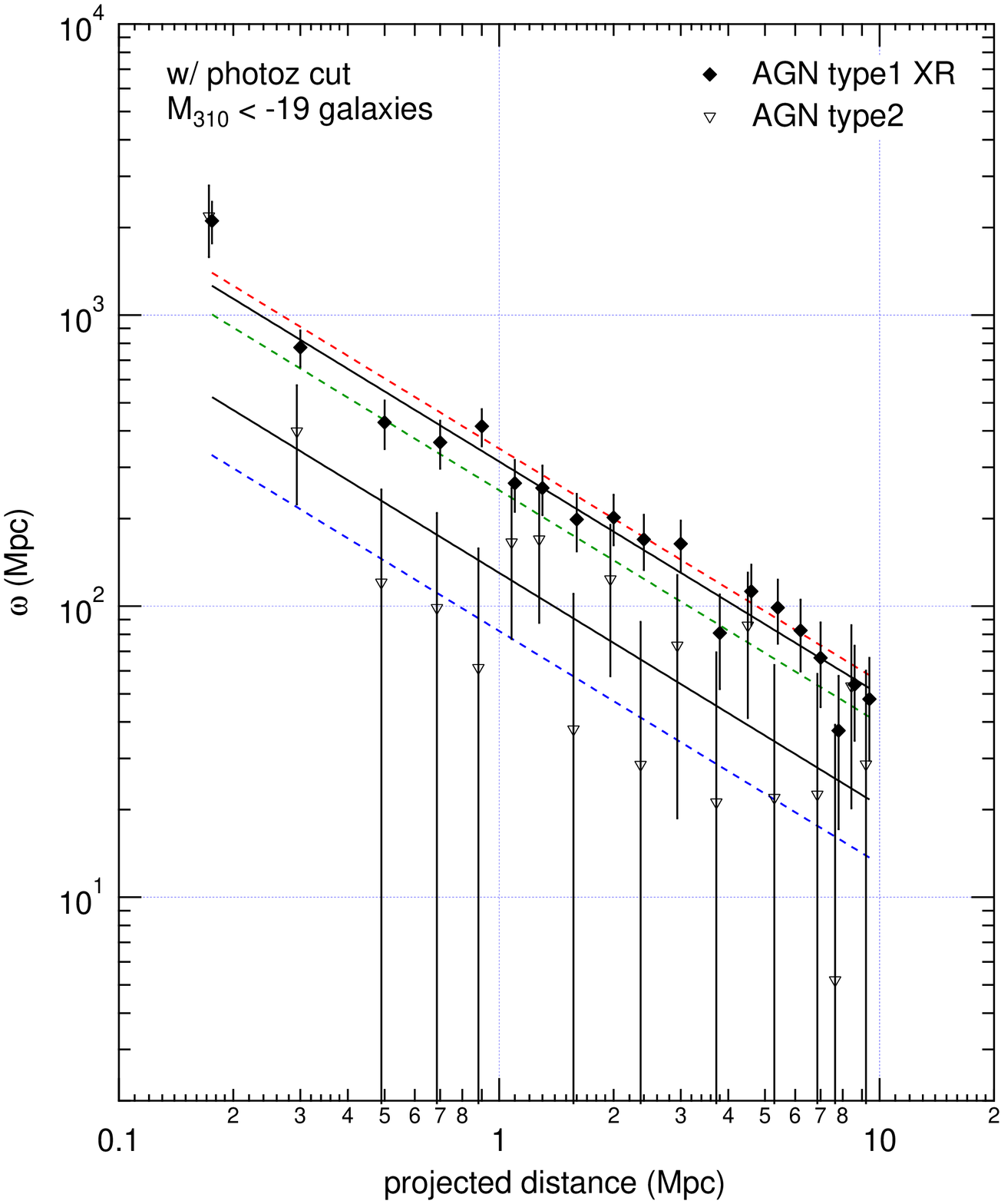}
  \end{center}
  \caption{Left: Projected cross-correlation functions derived for the 
    whole AGN,
    blue galaxy, and AGN-LG pair samples. The solid lines indicate the power
    law functions fitted to the data points. 
    The photo-$z$ selected galaxies were used and 
    the data were corrected for a reduction in the factor of 0.61.
    Right: The same plots as the left panel for AGN type~1 XR
    (X-ray and/or radio detection) and AGN type~2 samples. The solid lines
    are functions fitted to the data of these two samples, and the dashed lines are
    functions fitted to the whole AGN, AGN-LG, and blue galaxies, as shown in the left panel
    with the solid lines.
  }
    
  \label{fig:cc_galaxy}
\end{figure}

\begin{figure}
  \begin{center}
    \includegraphics[width=0.6\textwidth]{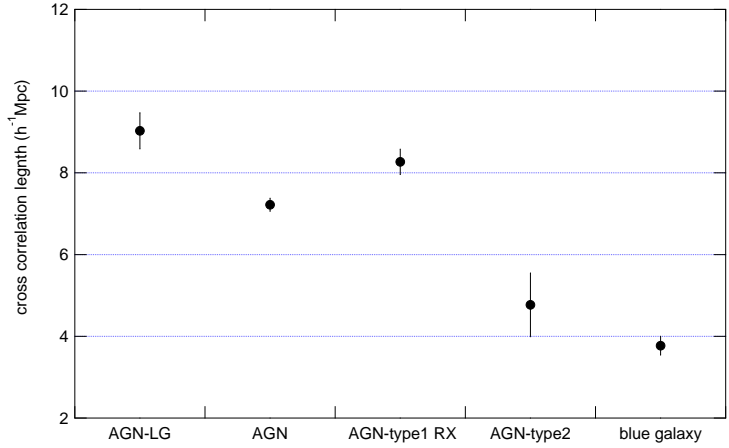}
  \end{center}
  \caption{Cross-correlation lengths measured for the five targets.}
  \label{fig:cc_length}
\end{figure}

The results of the fitting and fixed parameters are 
summarized in table~\ref{tab:fit_cc_hscsource}.
The cross-correlation functions obtained using the fitting parameter
are shown in the left panel of figure~\ref{fig:cc_galaxy} for the 
cases of the whole AGN, blue galaxy, and AGN-LG samples, and in the right panel
for the cases of AGN type~1 XR and AGN type~2 samples.
The cross-correlation lengths obtained for the five samples
are compared in figure~\ref{fig:cc_length}.
The cross-correlation lengths obtained for the whole AGN, AGN type1 XR, and
AGN-LG samples are significantly larger than those for the blue galaxy
and AGN type~2 samples.

The cross-correlation length for the AGN-LG sample is larger than that for
the whole AGN sample by four sigma, and is identical
to that for AGN type1 XR sample
within the margin of error.
The environment of AGN type~2 is similar to that of the blue galaxy, which
indicates that the AGNs of this sample are mostly caused by an internal
secular mechanism rather than an interaction with the external environment.

\begin{table}
  \tbl{Fitting parameters of cross-correlation functions between
    five types of target and HSC sources}{
    \begin{tabular}{crccccc}
\hline
      target type
      & $n^{\mathrm{a}}$
      & $\langle z \rangle^{\mathrm{b}}$
      & $r_0^{\mathrm{c}}$
      & $n_{\mathrm{bg}}^{\mathrm{d}}$
      & $\rho_{0}^{\mathrm{e}}$ 
      & $\gamma^{\mathrm{f}}$ \\
      & & & ${h^{-1}{\rm Mpc}}$ &   Mpc$^{-2}$ & $10^{-3}$Mpc$^{-3}$ & \\
\hline
AGN-LG pair   &   180 & 0.93 & 9.03 $\pm$ 0.44  & 4.807 $\pm$ 0.019 & 3.01 & 1.8 \\
     AGN      & 1 956 & 0.93 & 7.22 $\pm$ 0.16  & 4.695 $\pm$ 0.006 & 2.92 & 1.8 \\
blue galaxy   & 2 426 & 0.93 & 3.77 $\pm$ 0.27  & 4.747 $\pm$ 0.005 & 3.02 & 1.8 \\
AGN type-1 RX &   450 & 0.93 & 8.27 $\pm$ 0.31  & 4.718 $\pm$ 0.012 & 2.90 & 1.8 \\
AGN type-2    &   150 & 0.94 & 4.77 $\pm$ 0.78  & 4.890 $\pm$ 0.020 & 3.04 & 1.8 \\
\hline
    \end{tabular}}\label{tab:fit_cc_hscsource}
    \begin{tabnote}
$^{a}$number of target objects.
$^{b}$average redshift.
$^{c}$cross-correlation length and its error in one sigma.
$^{d}$average surface number density.
$^{e}$average space number density of galaxies at a redshift of the targets.
$^{f}$power index fixed to 1.8.
    \end{tabnote}
\end{table}

\subsection{Color distributions}

To investigate the difference in the composition of the galaxy types 
clustered around the three target types, namely, the whole
AGN, blue galaxy, and
AGN-LG pair, we derived the distribution of galaxy color within the
region of 0.2--2.0 Mpc from the targets.
For simplicity, in the subsequent analysis, excluding the average
number of clusters around the targets, the comparisons are made 
only for the three samples.

\begin{figure}
  \begin{center}
    \includegraphics[width=0.48\textwidth]{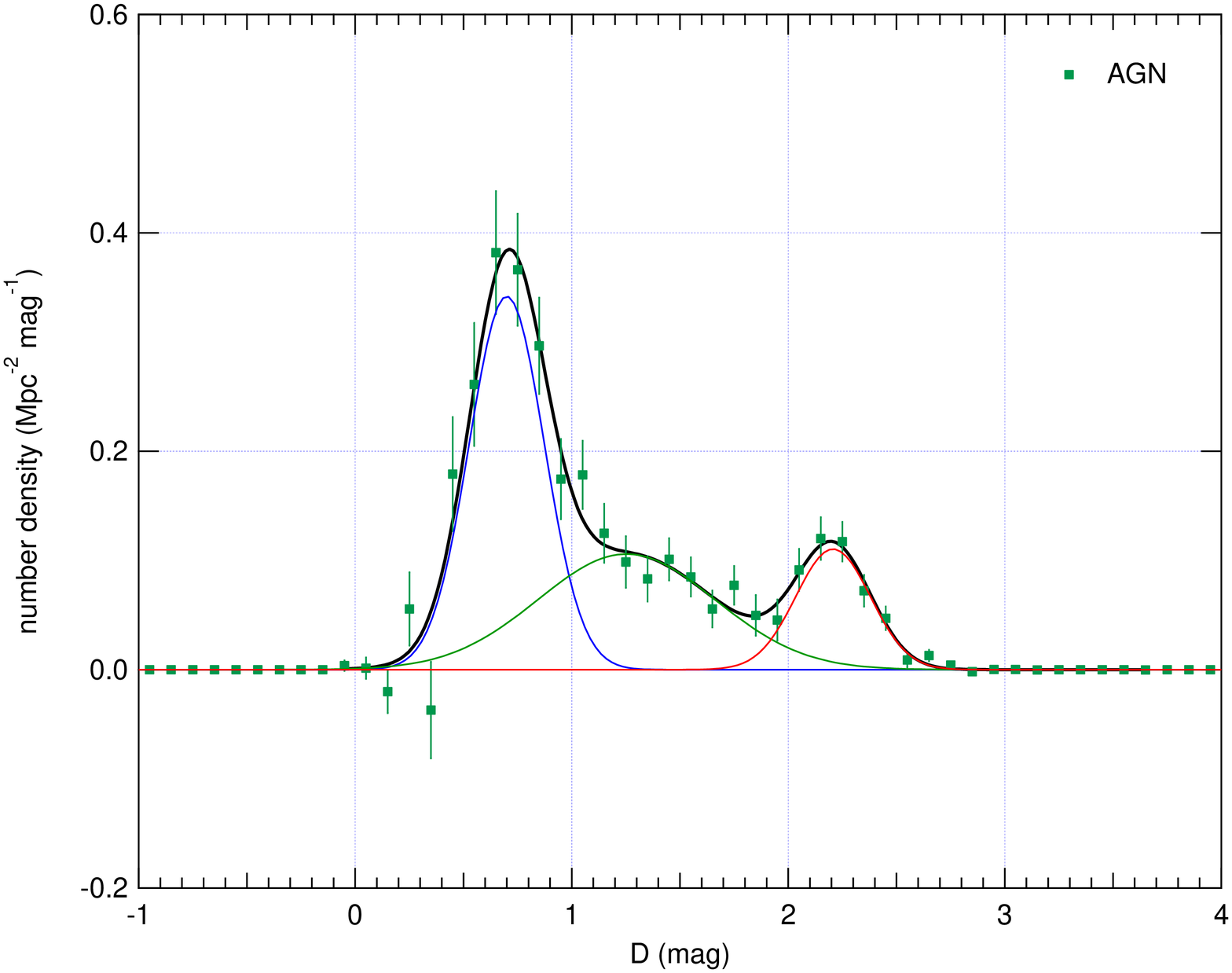}
    \includegraphics[width=0.48\textwidth]{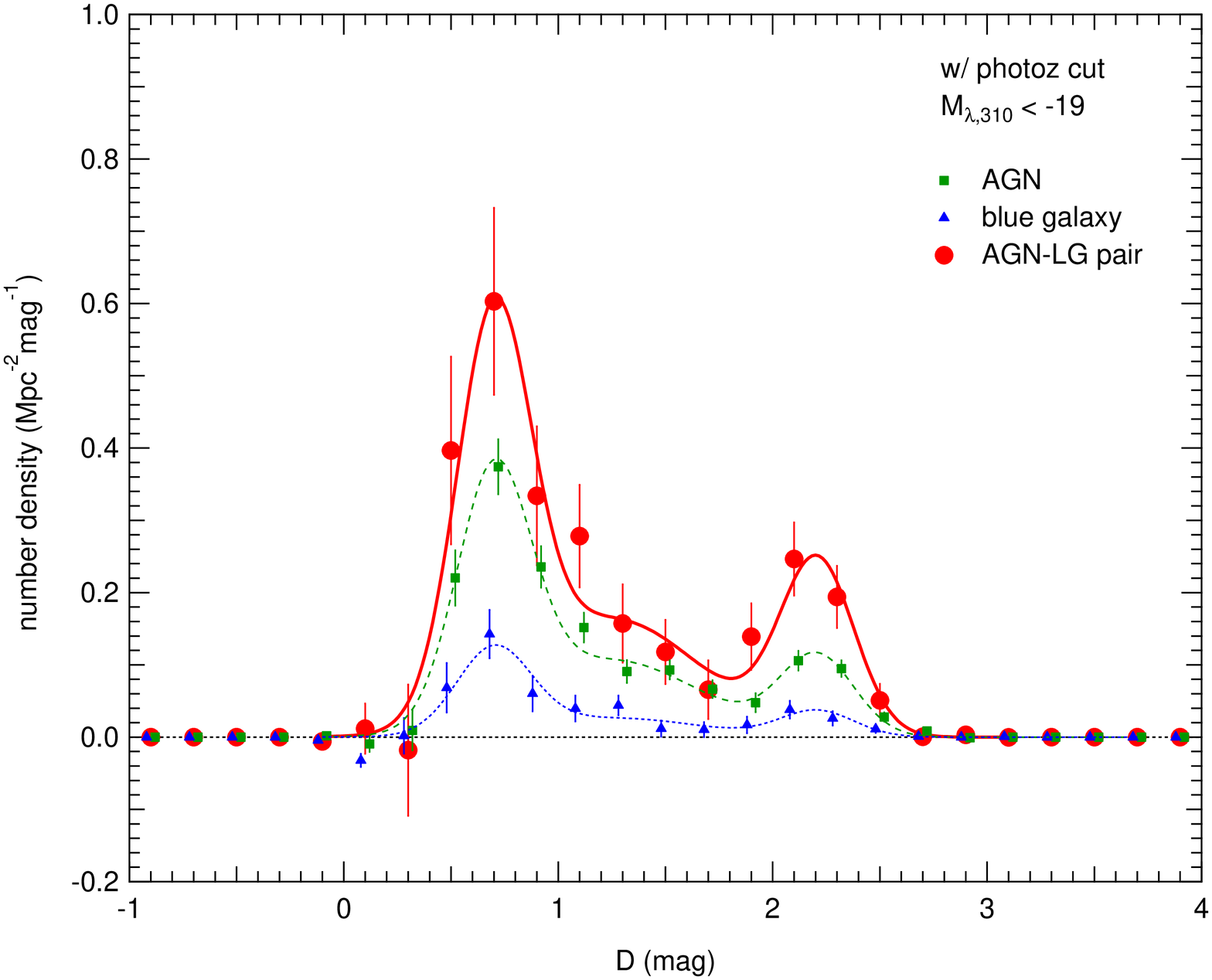}
  \end{center}
  \caption{Left: Color distributions of photo-$z$ selected galaxies around
    the AGNs of the whole AGN sample. The distributions were
    fitted using a linear combination of three Gaussian functions, each of which
    corresponds to blue-, green-, and red-type galaxies. 
    Right: Color distributions of photo-$z$ selected galaxies around
    AGNs or blue galaxies for the samples of whole AGN,
    blue galaxy, and AGN-LG pair.
    The distributions are fitted using three Gaussian functions.}
  \label{fig:dist_D}
\end{figure}

The color distribution was obtained by subtracting the color 
distribution in a lower
density region (7--9.8 Mpc) from that in a higher density region 
(0.2--2.0 Mpc).
The color of each HSC source was calculated according to the method
described in section~\ref{sec:dist_X}.
The distributions were derived from the HSC sources with a magnitude
brighter than
$M_{\lambda 320}$ 
=
$-19$~mag
and selected based on their photo-$z$.
The results are shown in figure~\ref{fig:dist_D}.

Each color distribution was fitted with a linear combination 
of three Gaussian distributions, each of which represents 
the distribution for
red sequence galaxies, blue cloud galaxies, and green valley
galaxies.
The fitting was first applied to the distributions for the
whole AGN sample (left panel of
figure~\ref{fig:dist_D}) by making all nine parameters free.
The fitting was then conducted by fixing the mean and standard deviation 
parameters of the three Gaussians to those determined with the fitting
to the whole AGN sample (right panel of figure~\ref{fig:dist_D}).
The fitted function is shown in the figure and the
values of the fitting parameters are summarized in table~\ref{tab:fit_D}.

\begin{figure}
  \begin{center}
    \includegraphics[width=0.48\textwidth]{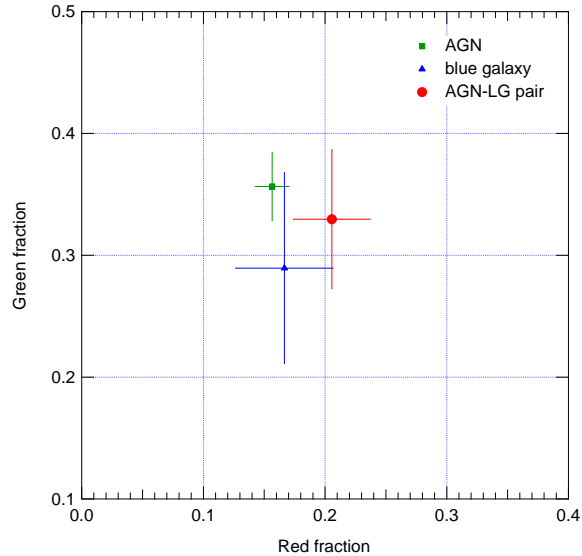}
  \end{center}
  \caption{Fractions of red and green galaxies obtained for the three samples.}
  \label{fig:RG_fraction}
\end{figure}

The number fractions of the green- and red-type galaxies 
to the total number of galaxies
for the three samples are plotted in figure~\ref{fig:RG_fraction}.
There is no significant difference among them.

\begin{table}
  \tbl{Fitting results for the color distributions.}{%
  \begin{tabular}{c ccc ccc ccc}
      \hline
 target type
 & $c$$^{\mathrm a}$
 & $\mu_{\mathrm B}$$^{\mathrm b}$
 & $\sigma_{\mathrm B}$$^{\mathrm c}$
 & $f_{\mathrm G}$$^{\mathrm d}$
 & $\mu_{\mathrm G}$$^{\mathrm b}$ 
 & $\sigma_{\mathrm G}$$^{\mathrm c}$ 
 & $f_{\mathrm R}$$^{\mathrm d}$
 & $\mu_{\mathrm R}$$^{\mathrm b}$ 
 & $\sigma_{\mathrm R}$$^{\mathrm c}$ \\
      \hline
AGN$^{e}$   & 0.304$\pm$0.016 & 0.70$\pm$0.02  & 0.17$\pm$0.03  
            & 0.36$\pm$0.13   & 1.25$\pm$0.16  & 0.41$\pm$0.13
            & 0.16$\pm$0.03   & 2.21$\pm$0.03  & 0.17$\pm$0.02 \\
AGN$^{f}$   & 0.304$\pm$0.014 & 0.70           & 0.17  
            & 0.36$\pm$0.03   & 1.25           & 0.41
            & 0.16$\pm$0.01   & 2.21           & 0.17  \\
AGN-LG pair$^{f}$ & 0.506$\pm$0.049 & 0.70          & 0.17
            & 0.33$\pm$0.06   & 1.25          & 0.41
            & 0.21$\pm$0.03   & 2.21          & 0.17 \\
blue galaxy$^{f}$ & 0.093$\pm$0.013 & 0.70          & 0.17
            & 0.29$\pm$0.08   & 1.25          & 0.41
  	    & 0.17$\pm$0.04   & 2.21          & 0.17 \\
      \hline
  \end{tabular}}\label{tab:fit_D}
  \begin{tabnote}
    $^{\rm a}$scaling factor of the Gaussian distribution for each galaxy component,
    $^{\rm b}$mean of the $D$ distribution for each component, and
    $^{\rm c}$standard deviation of the $D$ distribution for each component, in which the
    $^{\rm d}$fraction of green or red galaxy component.
    $^{\rm e}$fitting was applied by making all nine parameters free.
    $^{\rm f}$fitting was applied by fixing the mean and standard deviation parameters to the values
    obtained for the nine-parameter fitting to the data of the whole AGN sample.
  \end{tabnote}
\end{table}

\subsection{Absolute magnitude distribution}

To investigate the difference in the luminosity function of 
the galaxies clustered around the three target types, we derived the
absolute magnitude distributions within the region of 0.2--2.0 Mpc
from the targets, as applied for the color in the previous 
section.
They were derived separately for two galaxy types, namely, the blue 
and red galaxy types.
The blue types were selected by their color $D<1.4$, and the red types
were selected using $D\ge1.4$.
Because the color distribution of a green-type galaxy overlaps 
significantly with those of the red- and blue type-galaxies, we simply
divided the data into two galaxy types.
The absolute magnitude of each HSC source was calculated according 
to the method described in section~\ref{sec:dist_X}.

\begin{figure}
  \begin{center}
    \includegraphics[width=0.6\textwidth]{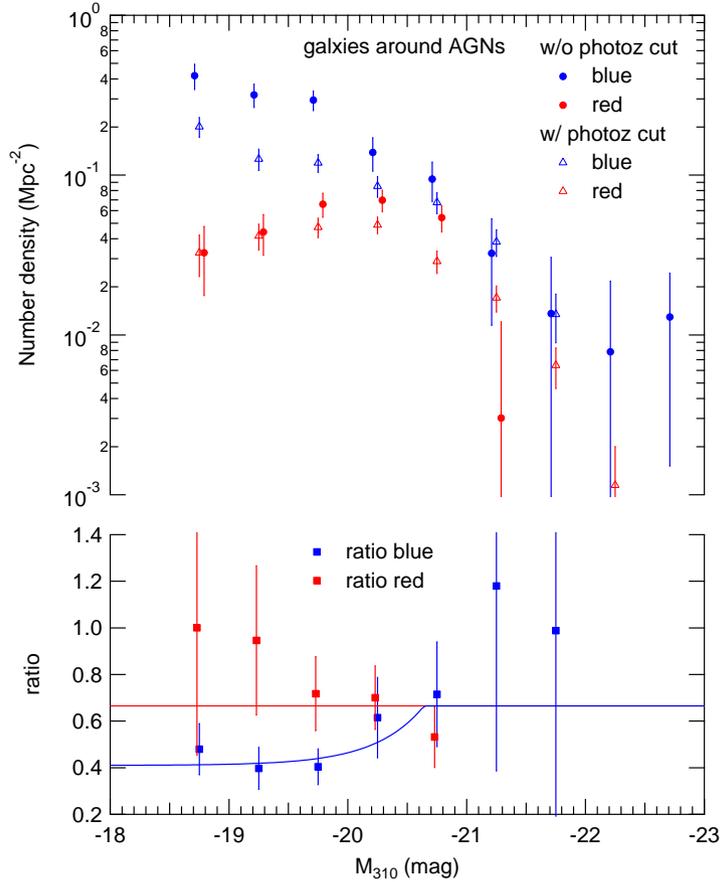}
  \end{center}
  \caption{
    Top: Absolute magnitude distributions of galaxies around AGNs for the
    whole AGN sample.
	The blue and red markers are distributions for the blue and red galaxies, respectively.
    The distributions shown with sold markers are derived from all galaxies,
    and those shown with open markers are from the photo-$z$ selected galaxies.
    Bottom: Ratios for the number of photo-$z$ selected galaxies against all galaxies.
	A color version is available on-line.}
  \label{fig:hist12_M_QSO}
\end{figure}

\begin{figure}
  \begin{center}
    \includegraphics[width=1\textwidth]{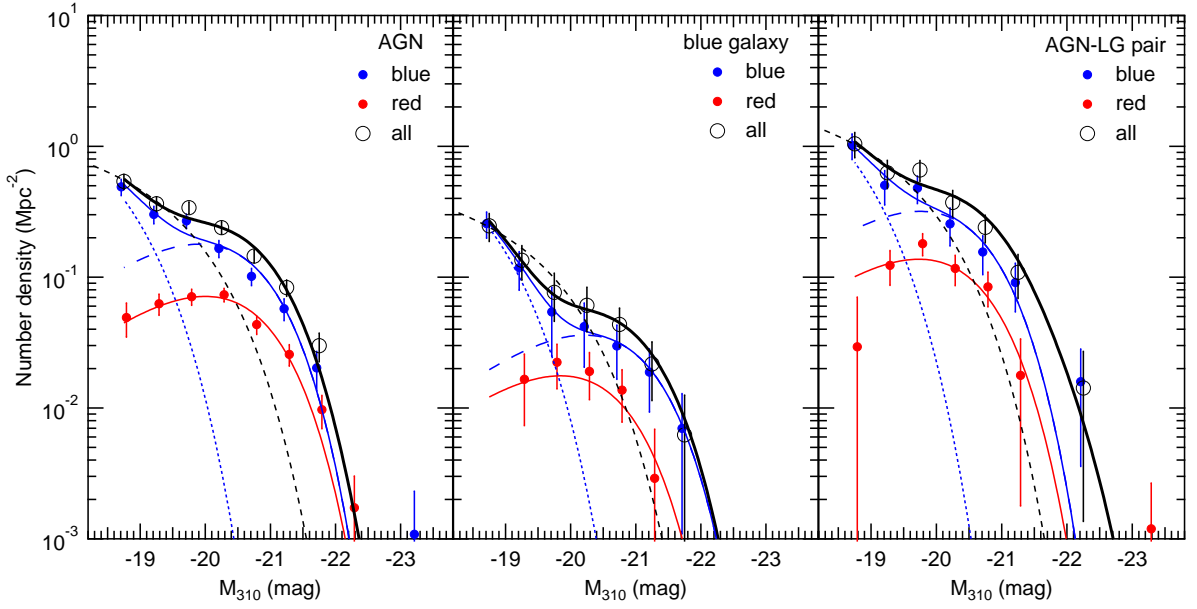}
  \end{center}
  \vspace{3em}
  \caption{Absolute magnitude distributions of photo-$z$ selected galaxies around
    target objects for
    three samples. The blue (red) closed circles indicate distributions for blue (red) galaxies,
    and the open circles are for all galaxies. The solid lines represent fitted functions
    expressed by a Schechter function (for red galaxies) or a combination of two
    Schechter functions (for blue galaxies). The thick lines represent sum of the functions
    for blue and red galaxies. 
    The dashed lines represent a model function derived from
    the luminosity functions described in the literature for $z = 0.95$, which are normalized to the data at
    $M_{\lambda 310} = -18.75$.
The dotted and long dashed lines represent fitted functions corresponding to the 
    primary and secondary components of blue galaxies, respectively.
A color version is available on-line.
}
  \label{fig:hist2_M}
\end{figure}

The absolute magnitude distributions derived were corrected for their
completeness
including the reduction from the photo-$z$ selection.
To estimate the reduction rate by the photo-$z$ selection as a function of
the absolute magnitude, we compared the
magnitude distribution obtained for all galaxies (without 
photo-$z$ selection) with that for the photo-${z}$ selected galaxies of the
whole AGN sample.
The results are shown in figure~\ref{fig:hist12_M_QSO}.
The top panel of the figure shows the comparison between the two galaxy 
samples, and the bottom panel shows the ratio
of the photo-$z$ selected galaxies to all galaxies.
We assumed that the reduction rate for the red galaxy type is constant for 
the entire range of magnitude.
The measured ratios are consistent with this assumption and the average ratio 
obtained was 0.665. 
In the case of the blue galaxy type, the rate of reduction decreases on the 
fainter side, as shown in the bottom panel of figure~\ref{fig:hist12_M_QSO}, and 
thus we interpolated the ratios using an analytic function, such that it increase 
to the average ratio given for the red galaxy type.
The interpolation is shown as a solid blue line in the same panel.
Using the reduction rate obtained in this way, we corrected the
magnitude of the distributions derived from the photo-$z$ selected galaxies for 
the three target types.
The results are shown in figure~\ref{fig:hist2_M}.
The plots are only shown in the magnitude range where the 
completeness
exceeds 50\%.

The magnitude distributions are fitted with a single Schechter
function~\citep{Schechter+76}
for the red galaxy type, whereas those for the blue galaxy type are
fitted with a combination of two Schechter functions with different 
parameters.
In fitting to the distributions for the red galaxy, we fixed the $\alpha$
parameter of the Schechter function to $\alpha=0$, which is the average
obtained for the three targets by making the $\alpha$ parameter free.
The reason of this is to compare $M_{*}$ among the three samples by 
fixing the $\alpha$ parameter to the same value.
The fitting parameters are summarized in table~\ref{tab:fit_mag_red}.

Looking at the plot for the blue galaxy sample (middle panel of 
figure~\ref{fig:hist2_M}),
the magnitude distribution for the blue galaxy type flattens at approximately
$M_{\lambda 310} \sim -20$ and steepens again at approximately $M_{\lambda 310} \sim -21$.
To reproduce this feature, we assumed two components,
one of which is characterized using the Schechter function with a larger
(fainter) characteristic magnitude $M_{*}$ and slope parameter $\alpha=-1.2$,
and the other is characterized with a smaller (brighter)
$M_{*}$ and flat slope parameter $\alpha=0$, which is the parameter used
for the red galaxy type.
In the cases of the whole AGN and AGN-LG samples, the secondary component with the brighter
$M_{*}$ parameter dominates over the primary component at magnitudes
$M_{\lambda 310} < -19$.
Because of that,
the parameter $M_{*}$ for the primary component was not well constrained, and thus the
$M_{*}$ parameter was fixed to the value obtained for the blue galaxy sample. 
In order to test the preference for adding the secondary component,
we also fit the absolute magnitude distribution with a single component 
model and calculated Akaike information criterion (AIC)~\citep{Akaike-74}
and Bayesian information criterion (BIC)~\citep{Schwarz-78} 
for both the single and two component models.
The AIC and BIC are information criteria to evaluate the goodness
of the statistical model from both the goodness of the fit and 
complexity of the model, and have been widely applied to astrophysics
problems~\citep[e.g.][]{Takeuchi-00,Liddle-07,Shirasaki+08}.

We used the following formula to calculate the AIC and BIC:
\begin{equation}
AIC = n \ln{\left(\frac{\chi^{2}}{n}\right)} + 2k,
\end{equation}
\begin{equation}
BIC = n \ln{\left(\frac{\chi^{2}}{n}\right)} + k\ln{(n)},
\end{equation}
where $\chi^{2}$, $n$, and $k$ are normalized residual sum of squares,
number of data points, and number of free parameters, respectively.
The result of the fitting parameters
and obtained AIC and BIC values
are summarized in table~\ref{tab:fit_mag_blue}.
The result for the two component model is shown in the first row
of each target type, and that for the single component
model is in the second row where the parameters corresponding to 
the second component are indicated with dashes.

According to the $\chi^{2}$ values, both models are acceptable 
in 90\% confidence level for all the samples.
If we compare the AIC and BIC values between the two models
for each sample, two component model is preferable for
the samples of blue galaxy and AGN-LG, and one component
model is preferable for the whole AGN sample.
This result support introducing the second component to the 
model as a plausible scenario especially for the case of blue galaxy sample.
If this is the case, it is natural to expect that two components
exist ubiquitously and there is difference in the mixing ratio 
depending on the environment.
The small difference in the AIC and BIC values for the
AGN and AGN-LG samples compared to the blue galaxy sample can 
be considered as a result of dominance of secondary component, 
which is inferred from the two-component fit, in the examined 
magnitude range.
In such a case AIC/BIC will preferentially select the single 
component model.

\begin{figure}
  \begin{center}
    \includegraphics[width=0.6\textwidth]{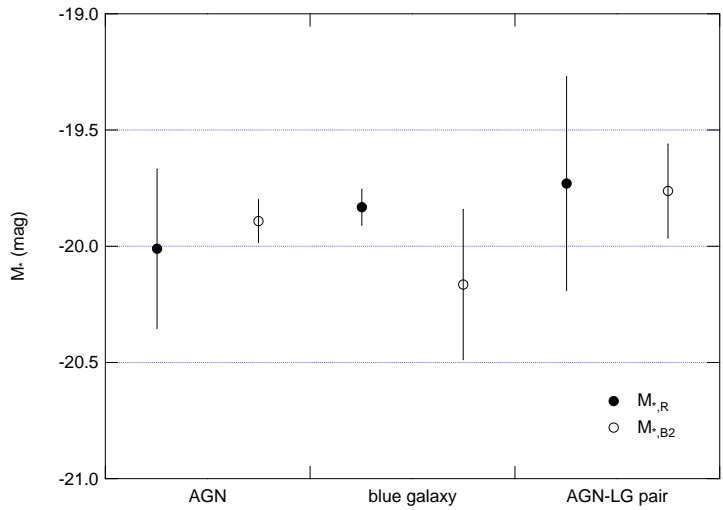}
  \end{center}
  \caption{$M_{*}$ parameters obtained for secondary component of blue galaxies
    ($M_{*,B2}$) and for red galaxies ($M_{*,R}$).}
  \label{fig:Mstar}
\end{figure}

The $M_{*}$ parameters obtained for the red-type galaxy and the secondary
component of the blue-type galaxy are shown in figure~\ref{fig:Mstar}.
As indicated in the figure, there are no significant differences in the
$M_{*}$ parameters among the three samples or between the
blue and red types.
Thus, the difference in the magnitude distribution among the three samples
are the fraction of the secondary component in the blue galaxy type
and the normalization factor of the luminosity function for each component.

\begin{table}
  \tbl{Fitting result for absolute magnitude distribution of red galaxies.}{%
  \begin{tabular}{c ccc cc}
      \hline
 target type
 & $\phi_{\mathrm{R}}$$^{\mathrm a}$
 & $\alpha_{\mathrm{R}}$$^{\mathrm b}$
 & $M_{*,\mathrm{R}}$$^{\mathrm c}$
 & $\chi^{2}$$^{\mathrm d}$
 & $n$$^{\mathrm e}$ \\
 \hline
 AGN &
 0.026$\pm$0.003 & 0.0 & $-$20.0$\pm$0.07 &
 3.59 & 7 \\
 blue galaxy &
 0.007$\pm$0.003 & 0.0 & $-$19.9$\pm$0.27 &
 5.54 & 7 \\
 AGN-LG pair &
 0.062$\pm$0.013 & 0.0 & $-$19.7$\pm$0.15 &
 5.53 & 7 \\
\hline
  \end{tabular}}\label{tab:fit_mag_red}
  \begin{tabnote}
$^{a}$~number density at $M_{\lambda 310} = -18$.
$^{b}$~$\alpha$ parameter of the Schechter function. This was fixed to 0.0.
$^{c}$~$M_{*}$ parameter of the Schechter function.
$^{d}$~normalized residual sum of squares,
$^{e}$~number of data points.
  \end{tabnote}
\end{table}

\begin{table}
  \tbl{Fitting result for absolute magnitude distribution of blue galaxies.}{%
  \begin{tabular}{c ccc ccc cccrr}
      \hline
 target type
 & $\phi_{\mathrm{B1}}$$^{\mathrm a}$
 & $\alpha_{\mathrm{B1}}$$^{\mathrm b}$
 & $M_{*,\mathrm{B1}}$$^{\mathrm c}$
 & $\phi_{\mathrm{B2}}$$^{\mathrm a}$
 & $\alpha_{\mathrm{B2}}$$^{\mathrm b}$ 
 & $M_{*,\mathrm{B2}}$$^{\mathrm c}$ 
 & $\chi^{2}$$^{\mathrm d}$ 
 & $n$$^{\mathrm e}$ 
 & $k$$^{\mathrm f}$ 
 & $AIC$$^{\mathrm g}$ 
 & $BIC$$^{\mathrm h}$ \\
 \hline
 AGN &
 0.96$\pm$0.17 & $-$1.2 & $-$18.3 &
 0.070$\pm$0.012 & 0.0 & $-$19.90$\pm$0.09 &
 6.50 & 7 & 3 & 5.5 & 5.3 \\
     &
 0.50$\pm$0.05 & $-1.2$ & $-$20.6$\pm$0.12 &
 --              & --     & --             &
 3.24 & 7 & 2 & $-1.4$ & $-1.5$ \\
 blue galaxy &
 0.56$\pm$0.29  & $-$1.2 & $-$18.3$\pm$0.56 &
 0.011$\pm$0.008 &  0.0 & $-$20.2$\pm$0.35 &
 0.05 & 7 & 4 & $-26.6$ & $-26.8$ \\
     &
 0.19$\pm$0.05 & $-1.2$ & $-$20.4$\pm$0.31 &
 --              & --     & --              &
 6.23 & 7 & 2 & 3.2 & 3.1 \\
 AGN-LG pair &
 1.84$\pm$0.58  & $-$1.2 & $-$18.3 &
 0.142$\pm$0.054 &  0.0 & $-$19.7$\pm$0.20 &
 2.13 & 7 & 3 & $-$2.3 & $-$2.5 \\
            &
 1.02$\pm$0.18  & $-1.2$ & $-$20.3$\pm$0.23 &
 --             & --     & -- &
 3.28 & 7 & 2 & $-$1.3 & $-$1.4 \\
\hline
  \end{tabular}}\label{tab:fit_mag_blue}
  \begin{tabnote}
$^{a}$~number density at $M_{\lambda 310} = -18$ for the primary (B1) and secondary (B2) components.
$^{b}$~$\alpha$ parameter of the Schechter function for the primary (B1) and secondary (B2) components.
they are fixed to $-$1.2 and 0.0, respectively.
$^{c}$~$M_{*}$ parameter of the Schechter function for the primary (B1) and secondary (B2) components.
$M_{*,B1}$ of AGN and AGN-LG pair are fixed to $-$18.3, which is the value obtained for blue galaxy.
$^{d}$~normalized residual sum of squares.
$^{e}$~number of data points.
$^{f}$~number of free parameters.
$^{g}$~Akaike information criterion.
$^{h}$~Bayesian information criterion.
  \end{tabnote}
\end{table}

\subsection{Stellar mass distribution}

\begin{figure}
  \begin{center}
    \includegraphics[width=0.5\textwidth]{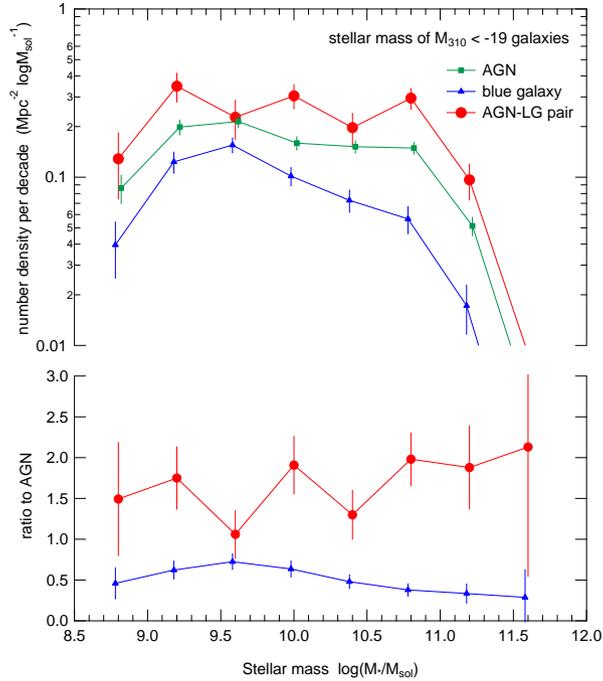}
  \end{center}
  \caption{Top: Stellar mass distribution of galaxies around target objects of
    three samples.
    Bottom: Ratios of the densities to those measured for
    the whole AGN sample.
  }
  \label{fig:hist_SM}
\end{figure}

Stellar mass distributions around the target objects for three samples were derived
using the photo-$z$ selected HSC sources with absolute magnitude of
$M_{\lambda 310} < -19$.
Figure~\ref{fig:hist_SM} shows a comparison between them.
The top panel shows the number densities at distances of 0.2--2.0 Mpc from the
target objects, which were obtained by subtracting the density distribution
at 7--9.8 Mpc from that at 0.2--2.0 Mpc.
The bottom panel shows the ratios to the number densities obtained 
for the whole AGN sample.

It can clearly be seen that HSC sources around the targets in
the whole AGN and AGN-LG 
samples have
a higher relative density than
those in the blue galaxy sample
at stellar masses of 
$M_{*} \ge 10^{10}M_{\solar}$.
The ratios for the
blue galaxy sample
shows a decreasing trend at above 10$^{9.6}M_{\solar}$.
There is no significant difference in the ratios for
the AGN-LG sample.

\subsection{positional distribution of clusters}

In previous sections, we compared the properties
of the environment for three samples
based on the properties of individual HSC sources, i.e.,
galaxies.
In this and the following sections, we investigate
the properties of their environment focusing on the clusters
of the HSC sources.

The method used to find clusters is described in
section~\ref{sec:cluster_finding}.
In creating the stellar mass or number density map, we used
photo-$z$ selected HSC sources with a magnitude of $M_{\lambda 310} < -19$~mag.
We selected clusters based on two density maps: one is a map of
the stellar mass density and the other is a map of the source number
density.

\begin{figure}
  \begin{center}
    \includegraphics[width=0.48\textwidth]{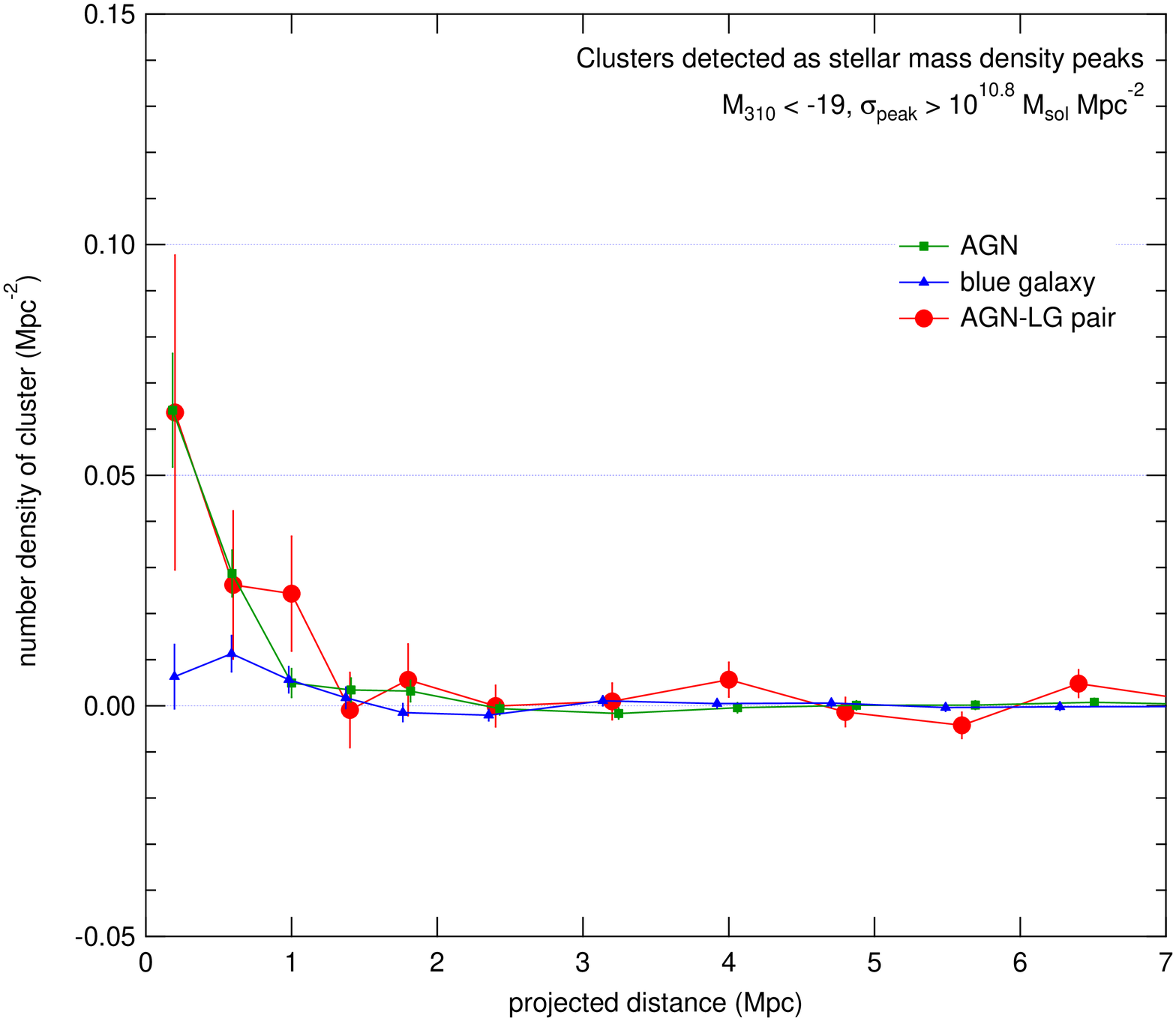}
    \includegraphics[width=0.48\textwidth]{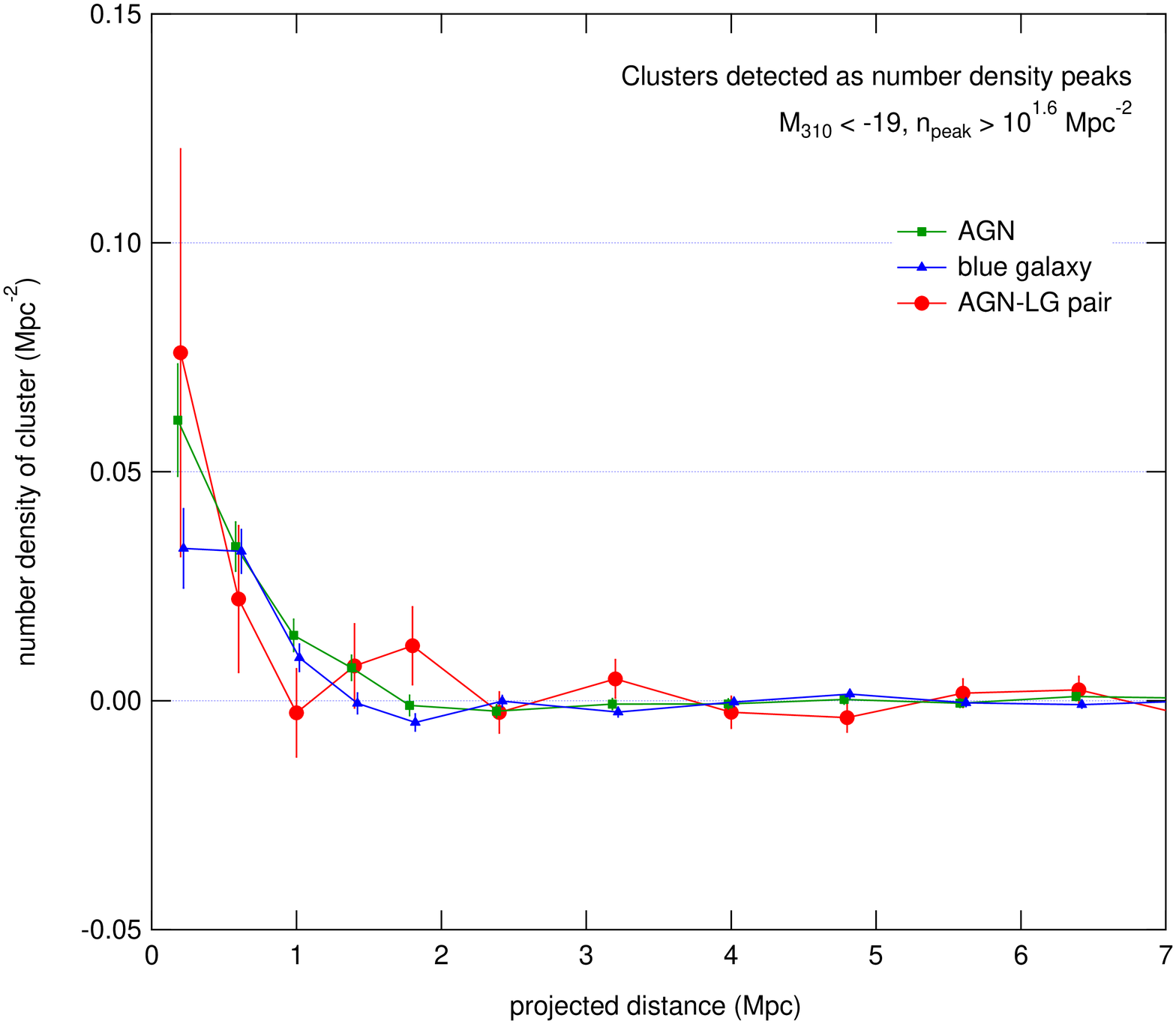}
  \end{center}
  \caption{
    Left: radial distributions of clusters detected as stellar
    mass density peaks for three samples. Offset (background)
    densities measured at 4--7 Mpc were subtracted.
    Right: Same plot as the left panel for
    clusters detected as the number density peaks.}
  \label{fig:cluster_density}
\end{figure}

Figure~\ref{fig:cluster_density} shows the radial number density
distributions of clusters found in the stellar mass density map 
(mass peak clusters, left-hand panel)
and clusters found in the number density map
(number peak clusters, right-hand panel).
The threshold for the counting cluster was set to peak densities of
$10^{10.8} M_{\solar}$Mpc$^{-2}$
and $10^{1.6}$Mpc$^{-2}$, respectively.
These numbers correspond to the detection threshold for clusters
by this method as will be shown in figure~\ref{fig:cluster_PK} of the 
next section.
The average cluster density at a projected distance of 4 to 7 Mpc
are subtracted from the density distribution.
In each panel, the distributions for the three samples are compared.
The uncertainties of the number density are derived based on the Poisson 
statistics and the error bars denote one sigma uncertainty.

The distributions of the mass peak clusters show an excess over the average at  $<$ 1.2~Mpc, and
the excess density increases toward the target objects
for the cases of the whole AGN and AGN-LG samples.
The significance of the excess is 7.1 and 2.9 sigma for the 
whole AGN and AGN-LG samples, 
respectively.
The excess is smaller and less significant (2.8 sigma) for the case of 
the blue galaxy sample.
The distributions of the number peak clusters show a significant excess 
at $<$ 1.2~Mpc distances for all three samples.
The significance of the excess is 6.8, 2.6, and 7.3 sigma for the 
whole AGN, 
AGN-LG, and blue galaxy samples, respectively.
The number density distributions are almost identical among the three
samples.
The average number of clusters that have a peak number density of
$>10^{1.6}$Mpc$^{-2}$ and are found at a distance of $<$ 1.2~Mpc from the 
target objects is 0.15.
Thus, for $\sim$85\% of the target objects,
clusters above the threshold
are unassociated with them.

To investigate whether isotropy occurs in the distribution of the clusters
in environments of AGN-LG pairs, we derived a density 
map for the distribution of the clusters in a reference frame defined by the
position of the AGN and LG.
The origin of the AGN-LG reference frame was set at the location of the AGN,
and the direction from the AGN to LG was defined as the $x$-axis direction.
The distance was then scaled such that the distance between the AGN and LG 
was normalized to five in the reference frame.
The $y$-axis was defined as the projected scaled-distance from the
AGN-LG axis.

\begin{figure}
  \begin{center}
    \includegraphics[width=0.6\textwidth]{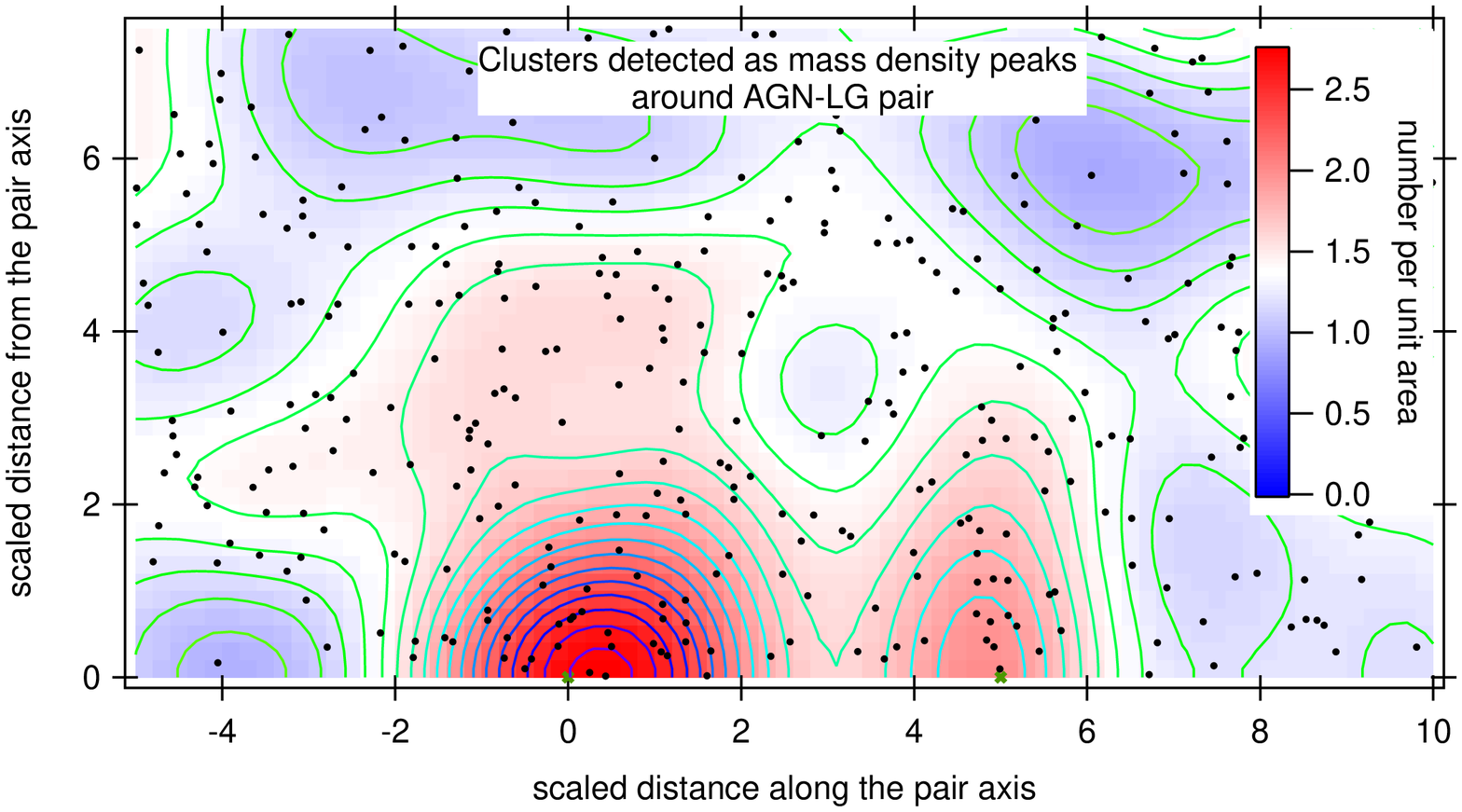}
    \includegraphics[width=0.6\textwidth]{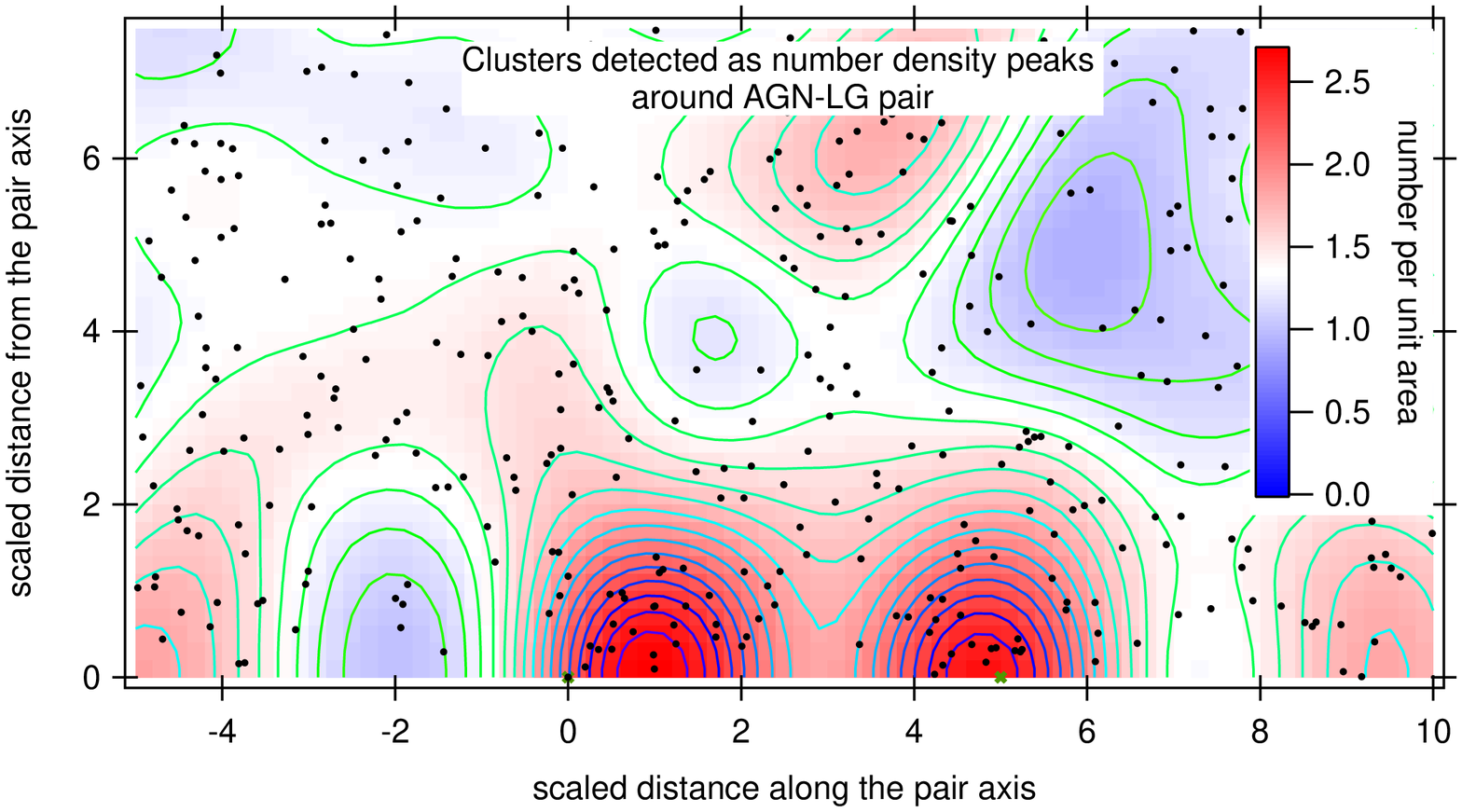}
  \end{center}
  \caption{Distributions of clusters in the AGN-LG frame. The top panel is for
    clusters found as the stellar mass density peaks, and the bottom panel is for
    clusters found as the number density peaks. 
    XY axes are scaled so that the
    the AGN and LG are 
    located at
    (0,0) and (5,0), respectively.
    The contours of the number density are drawn at a step 
    of 0.1 per unit area.
}
  \label{fig:cluster_AGN-LG-frame}
\end{figure}

By transforming the positions of all clusters to the AGN-LG reference
frame, they are plotted with solid circles in figure~\ref{fig:cluster_AGN-LG-frame}.
The top panel is for the mass peak clusters and
the bottom panel is for the number peak clusters.
The contours and color map were calculated by taking a convolution
with a 2D Gaussian with $\sigma=1$, as is applied when finding the
cluster peaks.

In both plots, a concentration of clusters around the AGN located 
at (0,0) and around the LG at (5,0) is clearly seen.
Looking at the positional distribution of the mass peak clusters
(top panel of the figure), a weak feature
elongated toward the vertical direction of the AGN-LG axis is indicated.

For the number peak cluster (bottom
panel of the figure), the elongation is not clear at the position of
LG ($x=5$) but is seen at the AGN position ($x=0$).
As a more outstanding feature, the peak of the concentration near 
the AGN is shifted toward the LG position.

To evaluate the significance of the anisotropy around the AGNs, we
carried out a Monte Carlo simulation for the distribution of clusters
in the AGN-LG frame.
The simulation was conducted for each dataset of AGN-LG pairs 
by assuming the power law plus constant
density distribution for clusters, as measured in the right panel of
figure~\ref{fig:cluster_density}.
The simulated positions of the clusters were converted into an AGN-LG 
frame according to the real positions of the AGN and LG.
The ratios of the number count at the LG side 
($x = 0$ -- $2$, $y < 2$) to the count at the anti-LG side ($x = -2$ -- $0$, $y < 2$) 
were then measured.
These numbers for the real observation are 26 at the LG side and 11 
at the anti-LG side, and thus the ratio is 0.70.

Among the 1000 sets of simulated samples, the maximum ratio was 0.61.
Comparing this value with the real observed value of 0.70, the 
probability of obtaining the observed anisotropy was estimated to be less than 0.1\%.

This anisotropy can be attributed to the overlap of clusters 
associated with the AGN and LG.
Such an offset is not significant in the distribution of clusters detected
as mass peaks, which may be due to a larger mass density for clusters
associated with AGNs than clusters associated with LGs.

\subsection{Peak density of clusters}

	To investigate the properties of the clusters associated with the
target objects, we derived the distribution of peak densities
for mass peak clusters and number peak clusters, in the same way as
conducted for the color, absolute magnitude, and stellar mass
described in the previous sections.
In deriving the distributions, a high-density region was taken
at $< 1$~Mpc and a low-density region was considered at 2--6~Mpc.

\begin{figure}
  \begin{center}
    \includegraphics[width=0.48\textwidth]{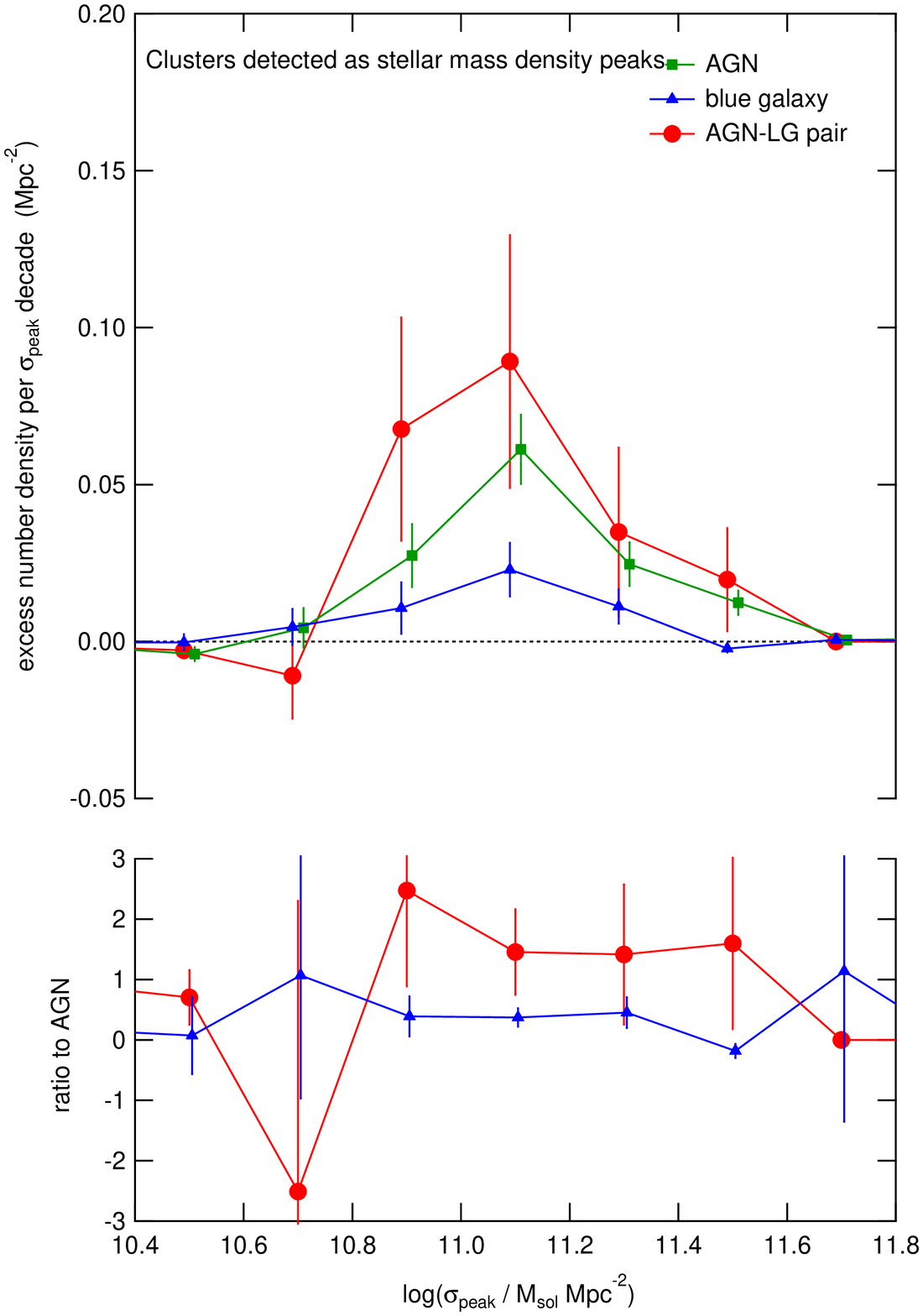}
    \includegraphics[width=0.48\textwidth]{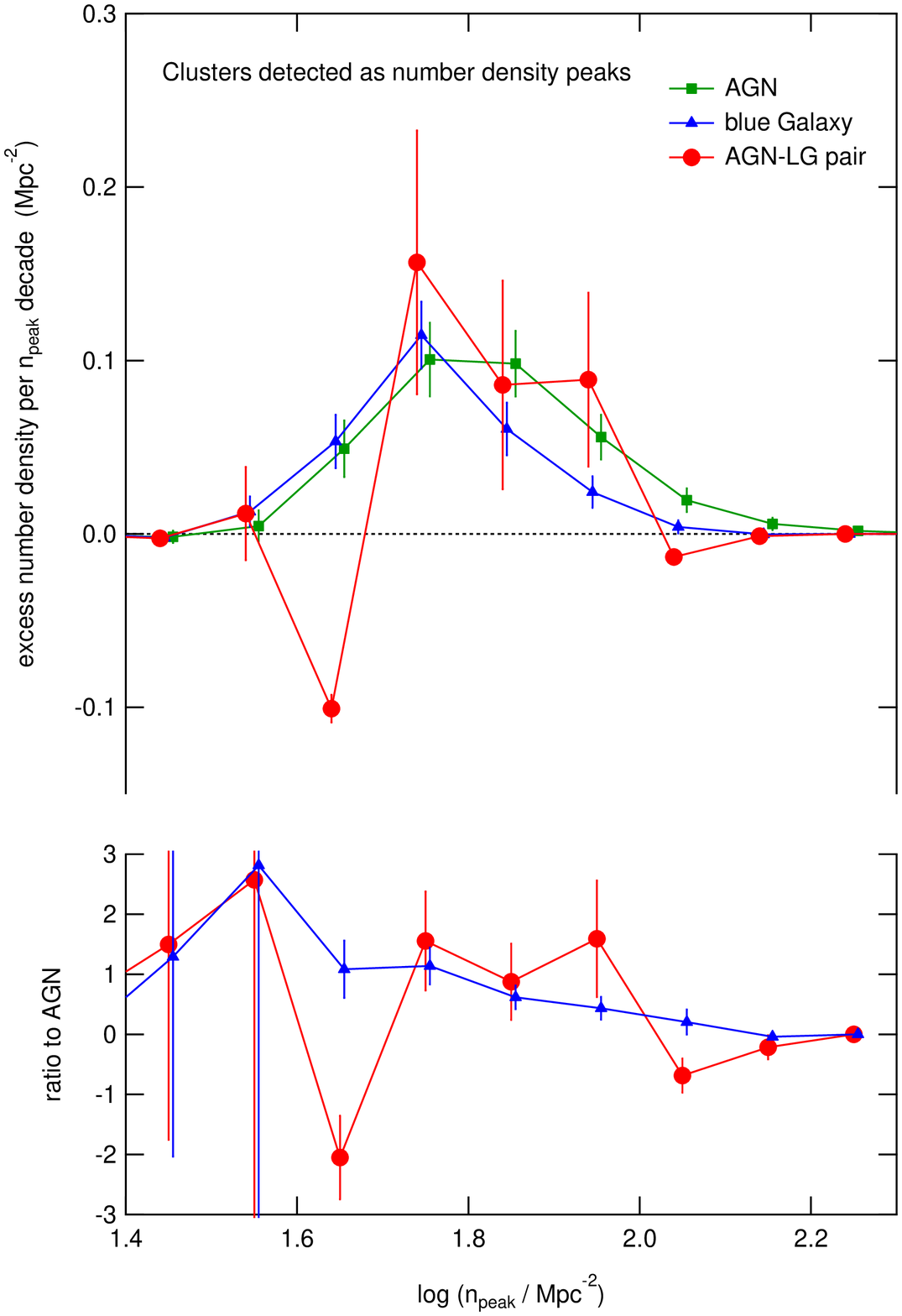}
  \end{center}
  \caption{Left: Distributions of peak stellar mass density for three
    samples. The ratios to the densities measured for the
    whole AGN sample are
    plotted in the bottom panel.
    Right: Same plot as the left panel for the peak number density.}
  \label{fig:cluster_PK}
\end{figure}

The left panel of figure~\ref{fig:cluster_PK} shows the distribution
of peak densities of the mass peak clusters, which demonstrates an excess of approximately
$\sigma_{\mathrm{peak}} = 10^{10.8}$--$10^{11.6}$M$_{\solar}$Mpc$^{-2}$.
The excess densities for the whole AGN and AGN-LG 
samples are larger 
than those for the blue galaxy sample at all peak densities.
No significant difference is seen in the excess densities
between the whole AGN and AGN-LG samples. 

The right panel of the figure~\ref{fig:cluster_PK} shows the
distribution of peak densities of the number peak clusters.
The excess densities for the whole AGN and blue galaxy samples
are almost identical at below $n_{\mathrm{peak}} < 10^{1.8}$Mpc$^{-2}$.
At higher peak densities, a decreasing trend is shown in
the ratios of density for the blue galaxy to whole AGN.
The excess densities for the AGN-LG sample are consistent 
with those of both the whole AGN and blue galaxy samples 
within the statistical error at $> 10^{1.7}$Mpc$^{-2}$.

A depletion at 10$^{1.6}$--10$^{1.7}$Mpc$^{-2}$ for the AGN-LG sample can be seen,
where no cluster was found at $<1$Mpc whereas
$\sim$5 
clusters were expected.
This may be due to the anisotropy of cluster distributions around the AGN of
the AGN-LG pairs, as the position of cluster is shifted toward the LG direction.

\subsection{Number distribution of cluster}

According to the results of the positional distribution of the clusters
in the AGN-LG frame, multiple clusters that are associated with 
the AGN and LG are expected to exist in the 
environment
near the AGN-LG pair.
To test this, we compared the number distribution of clusters found within a 
5~Mpc distance from the target objects.
The distance scale of 5~Mpc was chosen to count clusters located within 
1~Mpc from the LG, which is separated by 4~Mpc from the AGN at maximum.
The thresholds of the peak density were set to 
10$^{11}$$M_{\solar}$Mpc$^{-2}$ and 10$^{1.8}$Mpc$^{-2}$ for
the mass peak clusters and number peak clusters, respectively.
These thresholds approximately correspond to the
peak density of the distribution
as is shown in figure~\ref{fig:cluster_PK}.

\begin{figure}
  \begin{center}
    \includegraphics[width=0.48\textwidth]{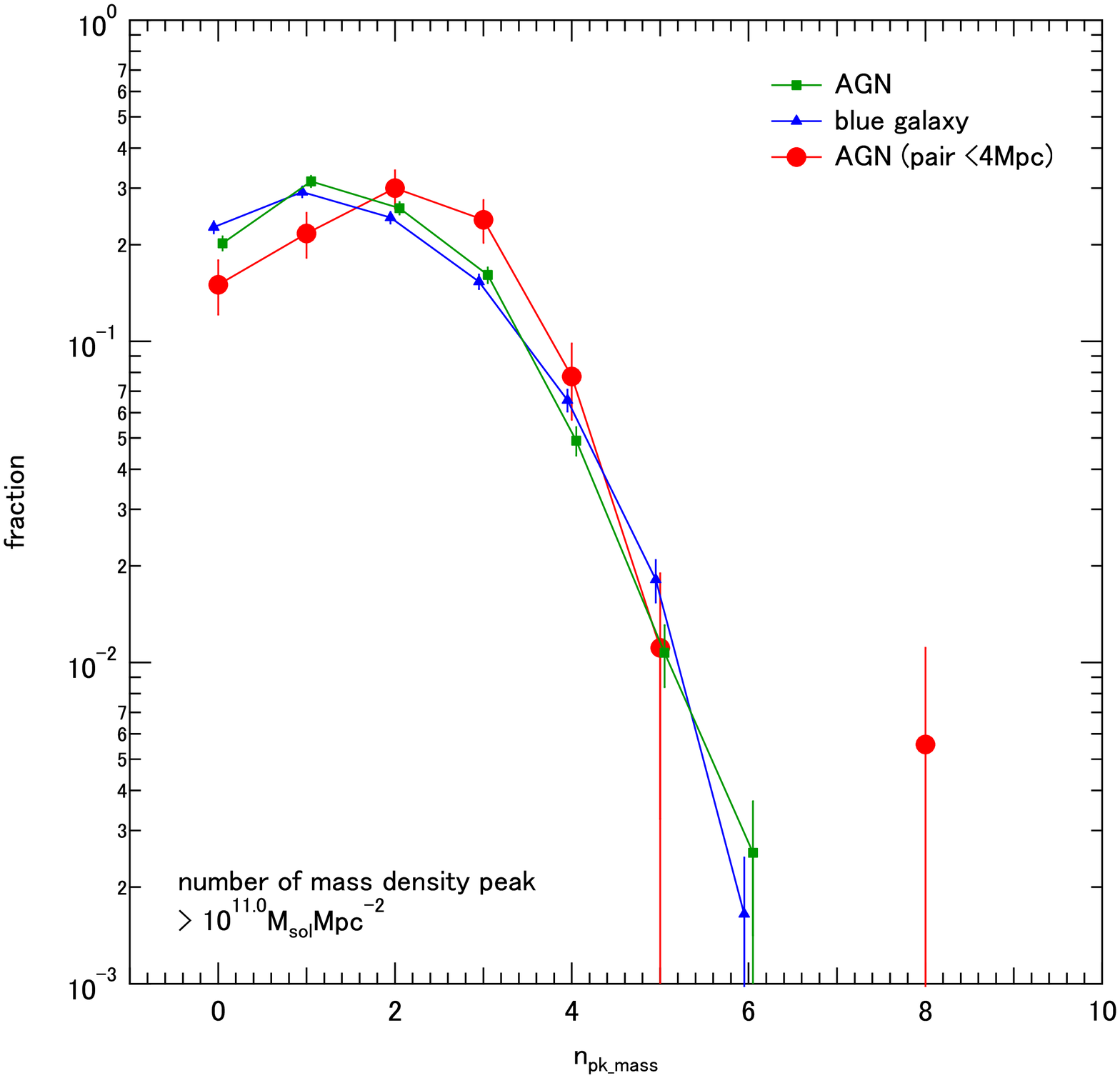}
    \includegraphics[width=0.48\textwidth]{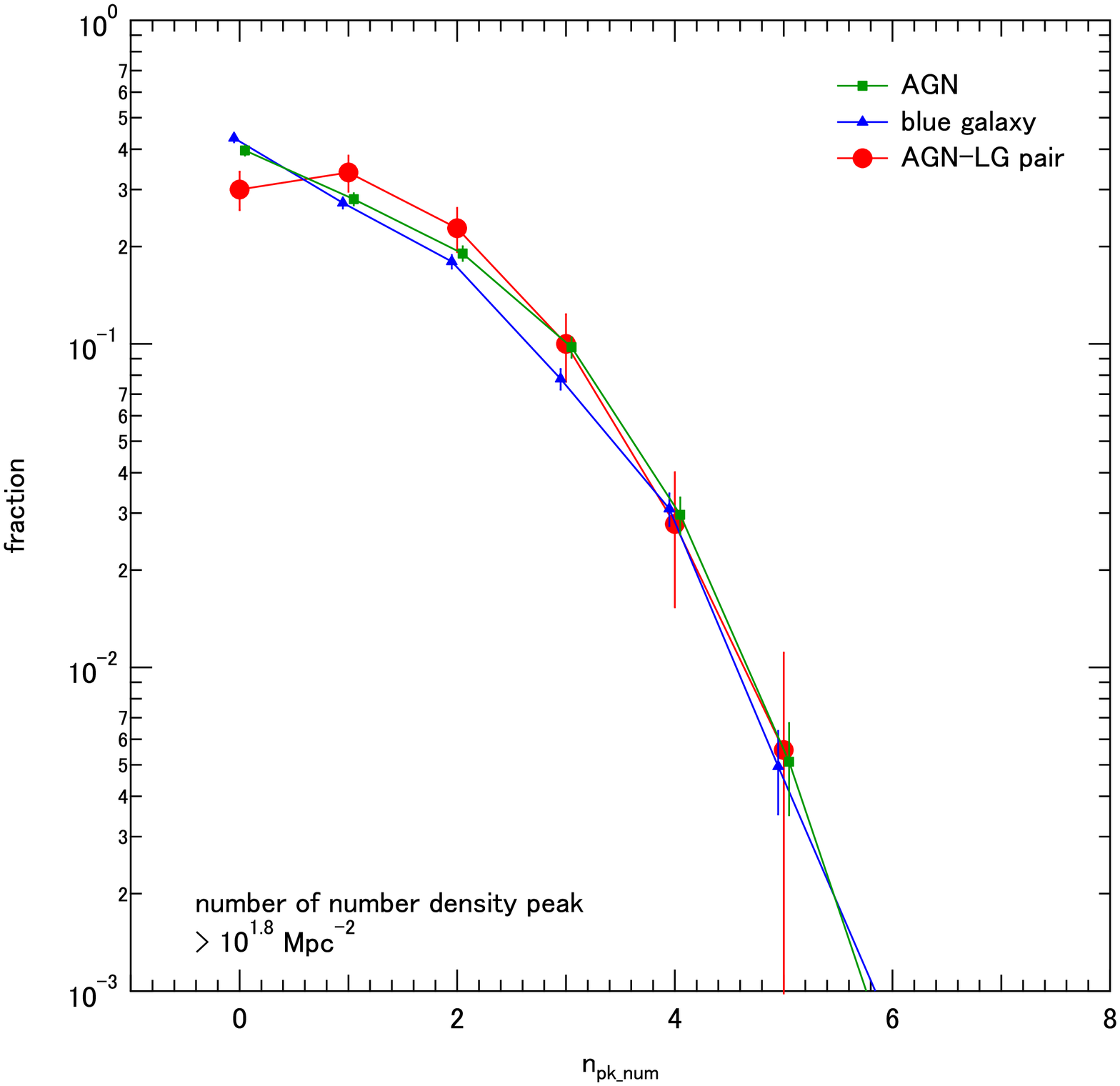}
  \end{center}
  \caption{Left: Normalized distribution of number of clusters detected
    as stellar mass density peaks and found at $<$5~Mpc from the target objects.
    The threshold peak density was set to 10$^{11} M_{\solar}$Mpc$^{-2}$.
    Right: Same plot for clusters detected as the number density peak. The
    threshold was set to 10$^{1.8}$Mpc$^{-2}$.
  }
  \label{fig:number_dist}
\end{figure}

Figure~\ref{fig:number_dist} shows the normalized distributions 
of the number of clusters for the three samples.
The left panel is for the mass peak clusters
and the right panel is for the number peak clusters.
In both cases, there is no significant difference between 
the distributions for the AGN and blue galaxy samples, whereas the distributions
for the AGN-LG sample are slightly shifted to a larger number.

\begin{figure}
  \begin{center}
    \includegraphics[width=0.6\textwidth]{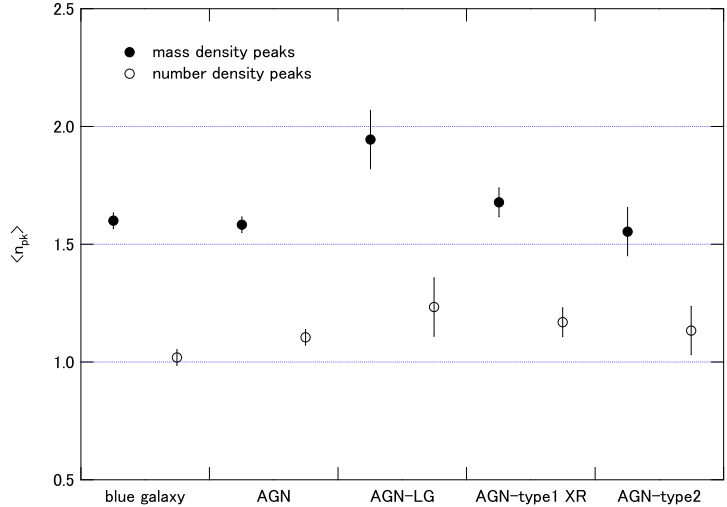}
  \end{center}
  \caption{Average number of clusters found at $<$5~Mpc from the target
    objects. Solid circles represent the number for clusters detected as
    stellar mass density peaks and open circles represent the number
    for clusters detected as number density peaks.}
  \label{fig:number_average}
\end{figure}

The average numbers of clusters per field are plotted in 
figure~\ref{fig:number_average} for the five samples, including
AGN type~1 XR and AGN type~2 samples.
We obtained $\langle n_{\mathrm{peak,mass}} \rangle$ 
($\langle n_{\mathrm{peak,num}} \rangle$) of
1.58$\pm$0.03 (1.10$\pm$0.03), 
1.60$\pm$0.03 (1.02$\pm$0.03), 
1.94$\pm$0.12 (1.23$\pm$0.09),
1.68$\pm$0.06 (1.17$\pm$0.05), and
1.55$\pm$0.10 (1.13$\pm$0.09)
for the whole AGN, blue galaxy, AGN-LG pairs, 
AGN type~1 XR, and AGN type~2, where 
$\langle n_{\mathrm{peak,mass}} \rangle$
($\langle n_{\mathrm{peak,num}} \rangle$)
represents the average number of mass peak clusters (number peak clusters).

The average number of mass peak clusters for the AGN-LG sample is
0.34 larger in $\sim$3 sigma than that of blue galaxy sample, 
whereas the difference in the average 
number of number peak clusters is small and less significant.
No significant difference can be seen among the other samples, which is
due to the dominance of the foreground and/or background clusters
unassociated with the targets.
These results indicate an environment in which the AGN-LG pairs has
multiple clusters with higher probability than the others has.

\section{Discussion}

\subsection{Comparison of galaxy bias with other measurements}

We measured the cross-correlation functions with the HSC sources, i.e.,
galaxies, for four types of AGN samples, i.e.,
whole AGNs with a mixture of 
any type, type~1 AGNs
detected in the X-ray and/or radio band (AGN type~1 XR), type~2 AGNs,
and AGN-LG pairs, as well as for a blue $M_{*}$ galaxy sample.
For comparison with other observations, we derived a linear bias
for galaxies clustering around those target objects.
We did not attempt to derive the linear bias for AGN itself because
of its difficulty arisen from the dominance of the evolved galaxies,
which were identified as a flat component
in an absolute magnitude distribution, in the AGN fields.

From our previous study\citep{Shirasaki+18}, it is known that 
there is strong correlation between the AGN and LGs
with 
magnitude brighter than $M_{*}$,
which is due to the evolution of $M_{*}$ into
the luminous side for galaxies around the AGNs.
In this study, we found that this evolution is related to the 
increase in the fraction of the secondary component in blue 
and red galaxies (referred to as a flat component collectively),
as shown in figure~\ref{fig:hist2_M}.
The dominance of the flat component makes it difficult to derive 
the absolute bias of the AGN from the cross-correlation with galaxies
because we need to consider the clustering
feature of the flat component.
It is also inadequate to assume a simple linear relation between 
the AGN-galaxy cross-correlation function and auto-correlation 
functions of the AGN and galaxy, considering that the spatial distribution 
of the AGNs and the flat component are presumably not independent 
of each other.

\begin{figure}
  \begin{center}
    \includegraphics[width=0.8\textwidth]{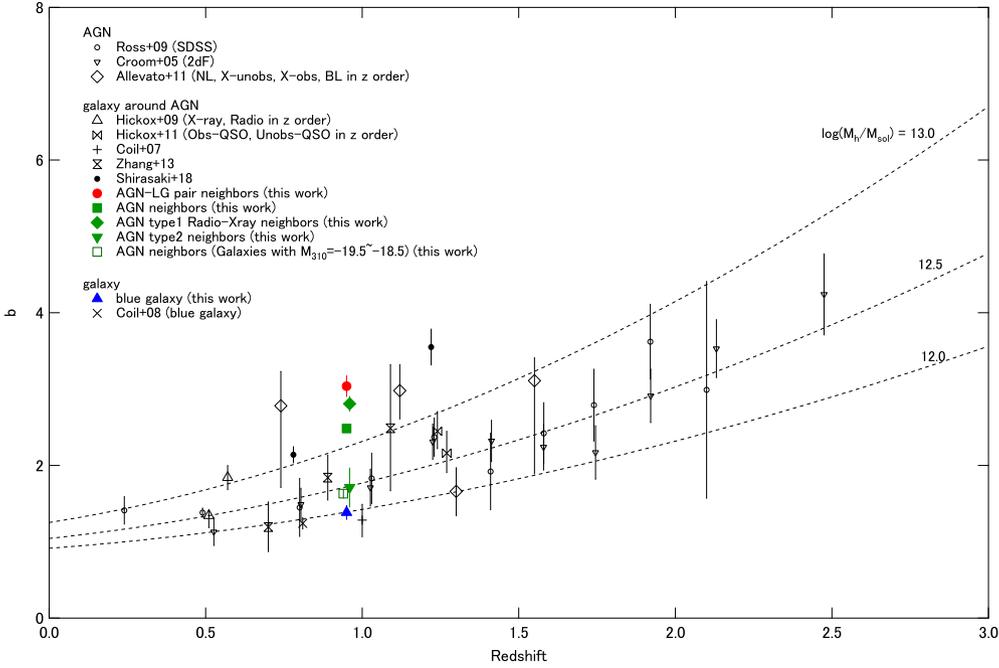}
  \end{center}
  \caption{
    Conditional absolute biases measured for galaxies
    ($M_{\lambda 310} < -19$) around
    five target types: AGN-LG pair, AGN (all types),
    AGN type~1 XR (X-ray and/or radio detection), AGN type~2
    and blue galaxies are plotted as solid makers.
    A bias calculated for galaxies with
    $M_{\lambda 310} = -19.5$--$-18.5$
    around the AGN (all types) is also plotted as an open square.
    The same bias corresponding to our previous results
    \citep{Shirasaki+18} and those obtained from the AGN-galaxy 
    cross-correlation by other authors 
    \citep{Coil+07,Hickox+09,Hickox+11,Zhang+13}
    are also shown.
 The absolute biases obtained for the AGN~\citep{Croom+05,Ross+09,Allevato+11} 
    and galaxy~\citep{Coil+08} auto-correlation are also plotted.
    The dashed lines represent a redshift evolution of bias for a DM 
    halo of the given mass, which is calculated in the same way as 
    described in \citet{Shirasaki+18}.
  }
  \label{fig:bias}
\end{figure}

For these reasons, we simply derived the linear bias 
$b_{\rm AG}$ for the galaxies 
around AGNs
from the cross-correlation length 
($r_{\rm AG}$)
and power index 
($\gamma$)
using the following formula
\citep{Shirasaki+18}:
\begin{equation}
b_{\rm AG} = \left( \frac{r_{\rm AG}}{8} \right)^{\gamma/2} J_{2}(\gamma)^{1/2}
             \left[ \sigma_{8} \frac{D(z)}{D(0)} \right]^{-1},
\label{eq:bias}
\end{equation}
where $J_{2}(\gamma)$ is defined as equation~(28) in \citet{Shirasaki+18}
and $D(z)$ is the linear growth factor given as equation~(30) in the same
literature.
The calculated biases for galaxies around the target AGNs and blue
galaxies are shown in figure~\ref{fig:bias} with solid markers, along
with those derived from other AGN or galaxy auto-correlation
\citep{Croom+05,Ross+09,Allevato+11,Coil+08} and AGN-galaxy
cross-correlation studies \citep{Coil+07,Hickox+09,Hickox+11,Zhang+13,Shirasaki+18}.
The biases found in the literature 
were calculated from the cross-correlation lengths and power index by
using 
equation~(\ref{eq:bias})
when they were derived from AGN-galaxy cross-correlation.
Because they were derived for galaxies around AGNs, we refer to the bias
as a conditional bias of galaxies.
The values derived from the AGN/galaxy auto-correlation were drawn 
from the literature as described.

In the same figure, the expected evolution in bias for DM haloes
of different masses are also shown as a reference.
We used the relation between dark matter halo mass and bias of the dark matter
halo derived by \citet{Sheth+01}.
As argued above, the conditional bias is affected by the evolution
of galaxies around the AGNs, and should not be used as an estimator
for the host DM halo mass.
Thus, the large biases obtained for our three AGN samples (
whole AGN, AGN type~1 XR, 
and AGN-LG pair) do not necessarily
mean that they are hosted by DM haloes with $>$10$^{13}M_{\solar}$.

As a reference, we also calculated the conditional bias for galaxies
with absolute magnitudes of $M_{\lambda 310} = -19.5$ -- $-18.5$,
where the contribution from the flat component becomes smaller
(but not negligible), for the whole AGN sample.
The corresponding cross-correlation length is 
$\sim$4.5~$h^{-1}$~Mpc.
This is shown in the same figure with an open square, and the bias
comes close to that obtained from the AGN auto-correlation~\citep{Croom+05,Ross+09}.
If we assume the auto-correlation length of galaxies 
with those magnitudes to be $\sim$3.8~$h^{-1}$Mpc, which is the value 
obtained for galaxies around the blue galaxy sample, the auto-correlation
length of the AGNs can be calculated as $r_{\rm AA} = r_{\rm AG}^2/r_{\rm GG}$ 
$\sim$5.4~$h^{-1}$Mpc assuming the same power index of $-$1.8 of correlation
function.
Then the bias is calculated as $\sim$1.9, which is consistent
with those obtained by \citet{Croom+05,Ross+09}.

The conditional biases calculated from our previous results~\citep{Shirasaki+18}
are also plotted in the same figure (solid black circles) for the $z0$ and $z1$
redshift groups.
The bias obtained for the whole AGN sample
in this work is consistent with
the increasing trend at higher redshifts.
The results obtained by \citet{Zhang+13}, who adapted an analysis method similar
to ours to the SDSS dataset, also show an increasing trend
of the conditional biases.
The conditional bias measured for the AGN-LG sample is the
largest of our samples, and is nearly the same as the one
obtained for the AGN type~1 XR sample within the margin of error.
The biases obtained for these two sub-types of
AGNs are larger than that obtained for the 
whole AGN sample,
which mostly consists of luminous type~1 AGN/QSO, by more than
three sigma.

The over density around AGNs with a radio emission
has been reported by numerous authors \citep[e.g.][]{Best+07,Hickox+09,Bradshaw+11}.
The clustering of X-ray selected AGNs is controversial.
\citet{Coil+09} 
found that X-ray AGNs are more clustered than optically selected
QSOs based on the cross-correlation with the same galaxy sample.
In contrast, \citet{Krumpe+12} reported that there is no
significant difference in the clustering of X-ray selected and 
optically selected broad-line AGNs.
Thus, there seems to be a variety in the clustering of the X-ray
selected AGNs, which is probably caused from the sample selection.
The X-ray selected AGNs in our AGN type~1 XR sample are likely to
reside in high density environment, as the sample
is dominated by X-ray selected AGNs as shown in table~\ref{tab:stat_agn}.

Our conditional bias for the blue galaxy sample is consistent with the
bias obtained by \citet{Coil+08}.
In the case of this sample, the contribution from the flat
component is almost negligible, and thus the bias obtained is expected to be
close to the halo bias.
The DM halo mass corresponding to the bias is $M_{h} \sim 10^{12} M_{\solar}$.
We found a smaller cross-correlation length for the type~2 AGN sample 
than those for the other AGN samples,
which is almost the same as that obtained for 
the blue galaxy sample.
The clustering of type~2 AGNs drawn from the SDSS DR4 was examined by
\citet{Li+06}, who reported that no significant difference
is shown between the type~2 AGNs and a reference sample of galaxies
on scales of larger than a few Mpc.
\citet{Allevato+11}, by contrast, reported a similar bias 
parameter for both broad-line and narrow-line AGNs selected in the
X-ray band.
Their measurements of bias are plotted in figure~\ref{fig:bias}, 
which shows a large bias for the narrow-line AGNs and a relatively 
smaller bias for X-ray obscured AGNs.
Considering the large error bar, it is difficult to draw a conclusive answer
to the difference in biases between the AGN types from their
measurements.

\citet{Hickox+11} measured the cross-correlation with galaxies for obscured
and unobscured QSOs.
  They argue that the obscured QSOs reside in denser environment than the
  unobscured QSOs in 1~sigma level.
  More significant evidence have been obtained by recent works
  \citep[e.g.][]{DiPompeo+14,DiPompeo+17}, whereas
  other authors argue that there is no significant difference between
  the environments of obscured and unobscured (or type~1 and type~2)
  QSOs~\citep[e.g.][]{Mendez+16}.
  Based on the clustering analysis for AGNs at the local Universe, 
  several authors have reported that obscured AGNs (or type~2 AGNs) 
  reside in denser environment than unobscured 
  (or type~1 AGNs)~\citep[e.g.][]{Powell+18,Krumpe+18,Jiang+16} if it
  is measured at small scales ($<$ 1~Mpc).
  Based on the HSC-SSP dataset, \citet{Toba+17} measured the clustering
  properties of infrared bright dust-obscured galaxies (DOGs), which
  is thought to be powered by active star formation and/or AGN hidden
  by surrounding dust, and they report large bias for those populations.

For comparison with our results,
conditional biases calculated from the 
cross-correlation lengths obtained by \citet{Hickox+11} for
obscured and unobscured QSOs
are plotted in the figure~\ref{fig:bias}.
Their biases are significantly smaller than our previous result 
obtained for whole AGNs at $z = 1.2$ 
and the extrapolation of this work for whole AGNs along
the bias evolution for the same halo mass, whereas the extrapolation of 
this work for type~2 AGNs is almost consistent with their results.
\citet{Coil+09} also reported a smaller bias for the AGN sample as compared
to our result for the whole
AGN sample at a redshift of $\sim 1$.
The calculated conditional bias for the AGN sample of \citet{Coil+09}
is almost the same as the bias obtained 
for a blue galaxy in \citet{Coil+08}.

It is possible that the inconsistency between their results and ours
comes from the difference in the estimation of the average number density of
the correlated galaxies.
In our analysis the galaxy density is estimated from the model of 
the luminosity function, which was derived from the luminosity functions
obtained by several deep surveys.
Because the statistics are usually low for galaxies at the luminous end
of the luminosity function derived from deep surveys, the flattening 
of the luminosity function at $<M_{*}$ owing to the emergence of the flat 
component, as observed in this study, could be completely missed.

This leads our analysis to an underestimation of the average number density
of galaxies, which results in a higher cross-correlation with bright galaxies.
In studies by \citet{Hickox+11} and \citet{Coil+07}, the average 
number density is estimated from the galaxy sample itself, and the sample
usually has sufficient statistics for detecting a flat component.
Despite the possible uncertainty in the estimation of the average number
density in our analysis, it is expected to be small considering the
consistency found for a blue galaxy sample between ours and \citet{Coil+08}.
Thus, the inconsistency could be due to the difference in the sample 
selection for the AGNs and/or galaxies.

Although there is a difficulty in comparing with other works
carried out using different methods and 
samples, a reliable comparison is possible if it
is made under the same method and same galaxy sample.
The differences found between our different target samples are 
more reliable, 
because they were compared under the same conditions to the greatest extent possible.
In the next section we discuss the properties of the environment
around AGN-LG pairs based on the results derived from fair comparisons.

\subsection{Properties of the environment around AGN-LG pairs}

As already discussed in the previous section,
the environment around AGN-LG pairs is characterized as a high-density
region if compared in a high luminosity range ($M< M_{*}$).
The projected cross-correlation function is as large as that of
AGNs with a radio and/or X-ray emission, which have been known to be located 
in a higher density region compared to the other type of AGNs.
As shown in figure~\ref{fig:hist_ratio_M310},
the luminosity of the AGNs in the AGN-LG pair sample is relatively
lower than those in the whole AGN sample.
One possible reason for this is that the AGNs in the AGN-LG pair
sample are dominated by AGNs obscured by their surrounding dust
to a certain degree.
There have been several studies suggesting that obscured AGNs tend
to reside in denser regions than unobscured
AGNs~\citep[e.g.][]{Hickox+11,DiPompeo+14,DiPompeo+17,Powell+18}.
Another possibility is the contribution of intrinsically less luminous 
AGNs, which are driven by, e.g., a quiescent accretion of hot halo 
gas~\citep{Keres+09,Fanidakis+13}.

We also found the excess in the average number of mass peak clusters 
for the AGN-LG sample
(figure~\ref{fig:number_average}) against those measured for 
the other samples.
In addition to this, we also found that the positional distribution of the 
number peak clusters departs 
from an isotropic distribution around each AGN and LG when they are
measured in the AGN-LG pair frame (bottom panel of figure~\ref{fig:cluster_AGN-LG-frame}).
These results indicate that, for a portion of the AGN-LG pairs,
at least two clusters are located around them.

The luminosity function of blue galaxies at distances of less than 2~Mpc 
from the AGNs of the AGN-LG pairs was measured, and it can be better
expressed with a linear combination of two Schechter functions 
with different $M_{*}$ and $\alpha$ values, as shown in 
figure~\ref{fig:hist2_M}.
The component characterized with a larger (fainter)
$M_{*}$ (primary component) is a main component at 
magnitudes fainter than $M_{\lambda 310} = -19$ mag.
The other component (secondary component), which dominates at 
magnitudes brighter than $M_{\lambda 310} = -20$ mag, is
characterized with a smaller (brighter) $M_{*}$ and 
a flat slope parameter of
$\alpha \sim 0$.

The luminosity function of red galaxies was also measured, and it 
is likely to have similar characteristics with the secondary component
found in the luminosity function of the blue galaxies.
Thus, both components are presumably produced by a common
mechanism, and we refer to them as a flat component hereafter.
The same features are also found for galaxies
in the other two 
samples, 
i.e., whole AGN and blue galaxy samples.

Comparing the ratios of the number density of the secondary component of
blue galaxies $\phi_{\mathrm{B2}}$ to that of the primary component 
$\phi_{\mathrm{B1}}$ (table~\ref{tab:fit_mag_blue}),
they are almost the same (0.078$\pm$0.014 for whole AGN 
and 0.075$\pm$0.023 for AGN-LG pair), whereas the fraction for the blue galaxy (0.021$\pm$0.018) 
is significantly smaller than that of the other two samples.

The comparison of the stellar mass distributions around the 
targets among the three samples
shows flatter distributions at larger stellar masses for the AGN and
AGN-LG than that for the blue galaxy (figure~\ref{fig:hist_SM}).
This can be explained by the higher fraction of the flat component,
which are typically luminous and thus have a larger
stellar mass, for the AGN and AGN-LG samples.

These results indicate that some type of AGN preferentially occurs in an
environment in which galaxies are rapidly evolving toward a red sequence.
The flat component of blue galaxies could be the intermediate state 
of a galaxy evolving toward a red sequence galaxy considering the similarity 
between the luminosity functions.

We were unable to find any 
significant difference in the properties of clusters for the AGN and AGN-LG 
pair samples within the immediate environment of $<$1 Mpc from the AGNs.
The clustering of clusters around the AGN are nearly the same for both 
samples (figure~\ref{fig:cluster_density}), and the distributions of 
peak density of the mass peak clusters and number peak clusters
are also identical within the 
statistical error (figure~\ref{fig:cluster_PK}).
A difference is seen if they are compared at a larger scale.

For the AGN-LG sample, we found that the average number of mass peak clusters
at $<$ 5~Mpc from the AGNs is larger than those for
the other samples (figure~\ref{fig:number_average}), which
indicates that the large clustering of LGs, 
i.e., the emergence of a flat component
for the AGN-LG pairs, is related with 
the larger clustering of DM haloes at a scale of several Mpc.
For the AGN type~1 XR sample, by contrast, we were unable to find
a significant difference in the number of mass peak clusters
from the other samples, i.e.,
whole AGN,
blue galaxy, and AGN type~2 samples.
This might be explained as two clusters being located too close
to be identified as separate mass peak clusters 
or already merged into a single large cluster
in the environment
of AGN type~1 XR.
At any rate, 
the existence of multiple clusters around AGN-LG pairs 
indicates that
there is some cluster-scale mechanism 
invoking AGN activity and evolution of surrounding
galaxies simultaneously.

  There is an argument that LGs are known to be highly clustered thus
  the environment of the AGN-LG pairs consequently should be 
  a high density region and have higher probability of association of
  nearby clusters.
  This argument, however, is not adequate in explaining why there is
  a large cross-correlation between AGN and LG seperated by several Mpc,
  which is an origianl question raised at the start of this work.
  We argue, instead, that the large-scale interaction of clusters
  could be a primary driver for producing AGNs and luminous galaxies
  in the restricted regions, which leads to the strong cross-correlation
  between them.
  Our observational results support this argument.
\section{Summary and conclusion}

We investigated the properties of the environment around pairs of
AGN and luminous galaxies (AGN-LG pairs) to understand
what causes the strong correlation between AGNs and LGs
found in~\citet{Shirasaki+18}.
From a comparison of the environmental properties measured for
four AGN samples, (whole AGN, AGN-LG, AGN type~1 XR, and AGN type~2 samples)
and one blue galaxy sample,
the following information was obtained:
\begin{enumerate}
\item AGNs are preferentially located at the environment where
  luminous galaxies ($M<M_{*}$) are enriched compared to the environment
  of blue $M_{*}$ galaxies. The measured cross-correlation lengths with
  galaxies ($M_{\lambda 310} < -19$) are 7.22$\pm$0.16~$h^{-1}$Mpc and 
  3.77$\pm$0.27~$h^{-1}$Mpc for 
  the whole AGN
  and blue galaxy samples, respectively. 
  All cross-correlation lengths are calculated for $\gamma = 1.8$.
\item The environment of type~2 AGNs is similar to that of
  blue $M_{*}$ galaxies. The measured cross-correlation length is
  4.77$\pm$0.78~$h^{-1}$Mpc for the AGN type~2 sample.
\item
  The luminosity of the AGNs in the AGN-LG sample
  is typically lower than
  that of the whole AGN sample,
  which may indicate that they are dominated by
  obscured and/or intrinsically less luminous AGNs.
\item AGNs in the AGN-LG sample
  are located in an environment of slightly higher density
  than those in the whole AGN sample.
  The measured cross-correlation length is
  9.03$\pm$0.44~$h^{-1}$Mpc for the AGN-LG sample.
  The cross-correlation is almost equivalent to that obtained
  for AGN type~1 (RX) sample, which is 8.27$\pm$0.31$h^{-1}$Mpc.
\item The cross-correlation length of AGNs with fainter galaxies comes close
  to that measured for blue galaxies. This indicates that the mass
  of the dark matter halo hosting AGNs is not particularly high but is 
  at most a few times
  as large as 
  that of the hosting blue galaxies.
\item The luminosity functions around the AGNs and blue galaxies are
  expressed by a linear combination of two Schechter functions: one
  represents the primary component at a faint end
  $M>M_{*}$, and the other one represents a flat component characterized by a flat slope parameter ($\alpha \sim 0$) and brighter
  $M_{*}$ than that of the primary component.
\item The ratio of the flat component to the primary component
  measured at $M_{\lambda 310} = -18$ is
  three-times higher in an AGN environment than in a blue galaxy environment.
  The larger cross-correlation lengths obtained for the AGN samples are
  mostly due to an enhancement of the flat component.
\item As expected from the enhancement of the flat component, which
  is characterized by bright, and hence a large stellar mass, 
  an enrichment of large stellar mass galaxies was measured around the AGNs
  as compared to around the blue galaxies.
\item The clustering of clusters detected as stellar mass density peaks
  (mass peak clusters)
  is larger around the AGNs than that around the blue galaxies, whereas the clustering
  of clusters detected as the number density peaks (number peak clusters)
  are almost the same between them.
  No statistically significant difference is shown between
  the whole AGN and
  AGN-LG samples.
\item The clustering of the mass peak cluster around the AGNs is almost equally
  larger at any peak density of clusters than that around the blue galaxies.
	A clustering of number peak clusters around the AGNs is similar for a smaller
  peak density ($<10^{1.8}$Mpc$^{-2}$), and becomes larger at a larger 
  peak density ($ \ge 10^{1.8}$Mpc$^{-2}$).
  No statistically significant difference is shown between
  the whole AGN and
  AGN-LG samples.
\item The anisotropic distribution of clusters was found in an
  environment of AGN-LG pairs if measured in the AGN-LG frame.
  We found that the peak position of the number peak clusters
  near the AGNs is shifted toward the LG direction, which indicates the
  coexistence of two clusters around the AGN-LG pairs.
\item An excess average number of mass peak clusters was found for
  the AGN-LG sample against the other samples, which again indicates
  the enhancement of the number of clusters in the AGN-LG pair 
  environments.
\end{enumerate}

Based on these findings, the following scenario regarding to the evolution 
of an AGN and a galaxy can be drawn:
AGNs, at least some classes of AGNs, are preferentially produced 
in an environment in which the number density of dark matter haloes is relatively high.
In such an environment, a star formation is ignited in multiple galaxies 
by a large-scale mechanism related with the multiple dark matter haloes, 
and then evolves through green valley galaxies with a quenched star formation,
finally becoming red sequence galaxies. 
An AGN is an episodic event produced along with a galaxy evolution, and thus
the strong correlation of an AGN and luminous galaxies is a natural consequence in
such an environment.
Collisions and/or interaction of dark matter haloes is a possible mechanism 
facilitating such a galaxy evolution at a large scale.

\begin{ack}

We would like to thank the anonymous referee for the constructive 
feedback, which helped us in improving the paper.
This work is based on data collected at the Subaru Telescope and 
retrieved from the HSC data archive system, which
is operated by Subaru Telescope and Astronomy Data Center,
National Astronomical Observatory of Japan.
The Hyper Suprime-Cam (HSC) collaboration includes the astronomical 
communities of Japan and Taiwan, and Princeton University.  The HSC 
instrumentation and software were developed by the National 
Astronomical Observatory of Japan (NAOJ), the Kavli Institute for 
the Physics and Mathematics of the Universe (Kavli IPMU), the 
University of Tokyo, the High Energy Accelerator Research 
Organization (KEK), the Academia Sinica Institute for Astronomy and 
Astrophysics in Taiwan (ASIAA), and Princeton University.  Funding
was contributed by the FIRST program from Japanese Cabinet Office, 
the Ministry of Education, Culture, Sports, Science and Technology 
(MEXT), the Japan Society for the Promotion of Science (JSPS),  
Japan Science and Technology Agency  (JST),  the Toray Science  
Foundation, NAOJ, Kavli IPMU, KEK, ASIAA,  and Princeton University.
This paper makes use of software developed for the Large Synoptic Survey
Telescope. We thank the LSST Project for making their code available as 
free software at 
http://dm.lsstorp.org.
The Pan-STARRS1 Surveys (PS1) have been made possible through contributions 
of the Institute for Astronomy, the University of Hawaii, the Pan-STARRS 
Project Office, the Max-Planck Society and its participating institutes, 
the Max Planck Institute for Astronomy, Heidelberg and the Max Planck 
Institute for Extraterrestrial Physics, Garching, The Johns Hopkins 
University, Durham University, the University of Edinburgh, Queen's 
University Belfast, the Harvard-Smithsonian Center for Astrophysics,
the Las Cumbres Observatory Global Telescope Network Incorporated, 
the National Central University of Taiwan, the Space Telescope Science
Institute, the National Aeronautics and Space Administration under Grant
No. NNX08AR22G issued through the Planetary Science Division of the NASA
Science Mission Directorate, the National Science Foundation under 
Grant No. AST-1238877, the University of Maryland, and Eotvos Lorand 
University (ELTE) and the Los Alamos National Laboratory.
Funding for the Sloan Digital Sky Survey IV has been provided by
the Alfred P. Sloan Foundation, the U.S. Department of Energy Office of
Science, and the Participating Institutions. SDSS-IV acknowledges
support and resources from the Center for High-Performance Computing at
the University of Utah. The SDSS web site is www.sdss.org.
SDSS-IV is managed by the Astrophysical Research Consortium for the 
Participating Institutions of the SDSS Collaboration including the 
Brazilian Participation Group, the Carnegie Institution for Science, 
Carnegie Mellon University, the Chilean Participation Group, 
the French Participation Group, Harvard-Smithsonian Center for Astrophysics, 
Instituto de Astrof\'isica de Canarias, The Johns Hopkins University, 
Kavli Institute for the Physics and Mathematics of the Universe (IPMU) / 
University of Tokyo, Lawrence Berkeley National Laboratory, 
Leibniz Institut f\"ur Astrophysik Potsdam (AIP),  
Max-Planck-Institut f\"ur Astronomie (MPIA Heidelberg), 
Max-Planck-Institut f\"ur Astrophysik (MPA Garching), 
Max-Planck-Institut f\"ur Extraterrestrische Physik (MPE), 
National Astronomical Observatories of China, New Mexico State University, 
New York University, University of Notre Dame, 
Observat\'ario Nacional / MCTI, The Ohio State University, 
Pennsylvania State University, Shanghai Astronomical Observatory, 
United Kingdom Participation Group,
Universidad Nacional Aut\'onoma de M\'exico, University of Arizona, 
University of Colorado Boulder, University of Oxford, University of Portsmouth, 
University of Utah, University of Virginia, University of Washington, University of Wisconsin, 
Vanderbilt University, and Yale University.
Funding for the DEEP2 Galaxy Redshift Survey has been
provided by NSF grants AST-95-09298, AST-0071048, AST-0507428, 
and AST-0507483 as well as NASA LTSA grant NNG04GC89G.
This research uses data from the VIMOS VLT Deep Survey, obtained from 
the VVDS database operated by Cesam, Laboratoire d'Astrophysique de 
Marseille, France. 
This paper uses data from the VIMOS Public Extragalactic Redshift Survey 
(VIPERS). VIPERS has been performed using the ESO Very Large Telescope,
under the "Large Programme" 182.A-0886. The participating institutions 
and funding agencies are listed at http://vipers.inaf.it 
Funding for PRIMUS is provided by NSF (AST-0607701, AST-0908246, AST-0908442,
AST-0908354) and NASA (Spitzer-1356708, 08-ADP08-0019, NNX09AC95G).
This research has made use of NASA's Astrophysics Data System.
This research has made use of the VO service toolkit developed by the Japanese
Virtual Observatory group at ADC, NAOJ.
We would like to thank to the members of HSC AGN WG for a fruitful discussion.
We would like to thank Editage (http://www.editage.com) for 
editing the first version of a manuscript for English language.
\end{ack}


\clearpage

\end{document}